\documentclass[aps, superscriptaddress]{revtex4-1}

\pdfoutput=1
\usepackage{graphicx}
\usepackage{amsmath}
\usepackage{amssymb}
\usepackage{bm}
\usepackage{color}

\newcommand{\mean}[1]{\langle{#1}\rangle}

\newcommand{\bra}[1]{\langle{#1}|}
\newcommand{\ket}[1]{|{#1}\rangle}

\newcommand{\Tr}{{\rm Tr}\hspace{0.07cm}}

\newcommand{\abs}[1]{{|#1|}}

\newcommand{\argmax}{\mathop{\rm arg~max}\limits}

\begin{document}
\title{Quantum Coherence Resonance}
\author{Yuzuru Kato}
\email{Corresponding author: kato.y.bg@m.titech.ac.jp}
\affiliation{Department of Systems and Control Engineering,
  Tokyo Institute of Technology, Tokyo 152-8552, Japan}

\author{Hiroya Nakao}
\affiliation{Department of Systems and Control Engineering,
  Tokyo Institute of Technology, Tokyo 152-8552, Japan}
\date{\today}

\begin{abstract}
It is shown that coherence resonance, 
a phenomenon in which regularity of noise-induced oscillations 
in nonlinear excitable systems is maximized at a certain optimal 
noise intensity, can be observed in quantum dissipative systems. 
We analyze a quantum van der Pol system subjected to squeezing,
which exhibits bistable excitability in the classical limit,
by numerical simulations of the quantum master equation.
We first demonstrate that quantum coherence resonance occurs in the semiclassical regime, namely, the regularity of the system's oscillatory response is
maximized at an optimal intensity of quantum fluctuations,
and interpret this phenomenon by analogy with classical noisy excitable systems using semiclassical stochastic differential equations.
This resonance persists under moderately strong quantum fluctuations for which the semiclassical description is invalid.
Moreover, we investigate even stronger quantum regimes and demonstrate that the regularity of the system's response can exhibit the second peak as the intensity of the quantum fluctuations is further increased.
We show that this second peak of resonance is a strong quantum effect that cannot be interpreted by a semiclassical picture, in which only a few energy states participate in the system dynamics.
\end{abstract}

\maketitle


\section{Introduction}
There are many real-world systems 
where noise brings order into their dynamics~\cite{horsthemke1984noise,schimansky1998noise, lindner2004effects,goldobin2005synchronization, teramae2004robustness, nakao2007noise, toral2001analytical, matsumoto1983noise}.
Stochastic resonance is a well-known example of such noise-induced order,
where the response of a system to a subthreshold periodic signal is
maximized at a certain noise intensity~\cite{gammaitoni1998stochastic}. 
It was first proposed as a model for the recurrence of the
ice ages~\cite{benzi1981mechanism, *benzi1982stochastic, nicolis1981stochastic}
and experimentally demonstrated using an ac-driven Schmitt trigger~\cite{fauve1983stochastic}.
Functional roles of the stochastic resonance in biological systems, such as those
in the mechanoreceptors of the crayfish~\cite{douglass1993noise} and in the
electrosensory plankton feeding of the paddlefish~\cite{russell1999use}, have also been revealed.
The possibility of stochastic resonance in quantum systems has also been considered theoretically~\cite{lofstedt1994quantum, grifoni1996coherent, grifoni1996quantum}
and the first experimental demonstration of the quantum stochastic resonance has recently been performed using an a.c.-driven single-electron quantum dot~\cite{wagner2019quantum}.

Coherence resonance, which was first coined by Pikovsky and Kurths~\cite{pikovsky1997coherence}, 
is another example of such noise-induced order, where regularity of noise-induced oscillations in an excitable system is maximized at a certain intermediate noise intensity.
It occurs as a result of two controversial effects of the noise, namely, increase in the regularity of the oscillatory response caused by noisy excitation and decrease in the regularity due to noisy disturbances.
Coherence resonance was first demonstrated near a saddle-node on invariant circle (SNIC) bifurcation~\cite{gang1993stochastic} and also near a supercritical Hopf bifurcation~\cite{pikovsky1997coherence} of limit cycles. Since then, a number of theoretical investigations have been carried out for various dynamical systems~\cite{lindner2004effects,anishchenko2007nonlinear}, including chaotic systems~\cite{palenzuela2001coherence}, spatially extended systems~\cite{perc2005spatial}, and realistic models of microscale devices such as semiconductor superlattices~\cite{mompo2018coherence} and optomechanical systems~\cite{yu2018noise}.
It has also been used to model the periodic calcium release from the endoplasmic reticulum in a living cell~\cite{shuai2002optimal, meinhold2002analytic}.
Experimental demonstration of coherence resonance has been performed in electrical circuits~\cite{postnov1999experimental},
lasers~\cite{giacomelli2000experimental,ushakov2005coherence}, chemical reactions~\cite{zhou2002experimental}, 
optically trapped atoms~\cite{wilkowski2000instabilities},
carbon nanotube ion channels~\cite{lee2010coherence}, 
and semiconductor superlattices~\cite{shao2018fast}.

In contrast to stochastic resonance, coherence resonance in quantum systems has not been explicitly discussed in the literature.
In lasers, the noise essentially comes from quantum mechanical effects~\cite{hempstead1967classical,lindner2004effects}, but coherence resonance has so far been analyzed only from a classical viewpoint.
Considering the recent developments in the analysis of limit-cycle oscillations in quantum dissipative systems where synchronization phenomena similar to those in noisy classical oscillators are observed~\cite{lee2013quantum, walter2014quantum, sonar2018squeezing, weiss2017quantum, kato2019semiclassical, kato2020semiclassical, lorch2016genuine},
it is natural to analyze coherence resonance in quantum dissipative systems.

In this paper, we demonstrate that coherence resonance occurs 
in a simple quantum dissipative system, which we call \textit{quantum coherence resonance}.
We analyze a quantum van der Pol (vdP) system~\cite{lee2013quantum, walter2014quantum, sonar2018squeezing, weiss2017quantum, kato2019semiclassical, kato2020semiclassical, lorch2016genuine, bastidas2015quantum, es2020synchronization, bandyopadhyay2020quantum, mok2020synchronization}
subjected to squeezing~\cite{sonar2018squeezing},
which is near a SNIC bifurcation 
\cite{strogatz1994nonlinear,guckenheimer1983nonlinear}
of a limit cycle in the classical limit, in the semiclassical and strong quantum regimes by direct numerical simulations of the quantum master equation. 
In the semiclassical regime, we show that the \textit{normalized degree of coherence}, which characterizes regularity of the system's oscillatory response,
is maximized at a certain optimal intensity of quantum fluctuations, and discuss this resonance phenomenon on the analogy of classical noisy oscillators by using a stochastic differential equation (SDE) for the system state in the phase space fluctuating along a deterministic classical trajectory due to small quantum noise.
We show that this peak in the degree of coherence persists even in the quantum regime where the semiclassical SDE is not valid.
We then consider even stronger quantum regimes and show that the system can exhibit the second peak in the degree of coherence when the intensity of quantum fluctuations is further increased.
We argue that this second peak of resonance is an explicit quantum effect resulting from small numbers of energy states participating in the system dynamics, which cannot be described using a semiclassical picture.


\section{Quantum van der Pol system subjected to squeezing}
As a minimum model exhibiting quantum coherence resonance, 
we consider a single-mode quantum vdP model subjected to squeezing~\cite{sonar2018squeezing}, 
which is an excitable bistable system slightly before the onset 
of spontaneous limit-cycle oscillations in the classical limit. 
We consider the case where the squeezing is generated by a degenerate parametric amplifier~\cite{gardiner1991quantum}. 
Such a system can be experimentally implemented using trapped ions and optomechanics as discussed in Ref.~\cite{sonar2018squeezing}.

We denote by $\omega_{0}$ the frequency parameter of the vdP system, which gives the frequency of the harmonic oscillation when the damping and squeezing are absent, and by $\omega_{sq}$ the frequency of the pump beam of squeezing. 
In the rotating coordinate frame of frequency $\omega_{sq}/2$, 
the evolution of the system is described by the following quantum master equation
\cite{sonar2018squeezing, kato2019semiclassical}:
\begin{align}
  \label{eq:qvdp_me}
  \dot{\rho} = 
  -i \left[  - \Delta a^{\dag}a 
    + i \eta ( a^2 e^{-i \theta} - a^{\dag 2} e^{ i \theta}  )
    , \rho \right]
  + \gamma_{1} \mathcal{D}[a^{\dag}]\rho + \gamma_{2}\mathcal{D}[a^{2}]\rho,
\end{align}
where $\rho$ is the density matrix of the system,
$a$ is the annihilation operator that subtracts a photon from the system,
$a^{\dag}$ is the creation operator that adds a photon to the system,
$\Delta = \omega_{sq}/2 - \omega_{0}$ is the detuning of
the half frequency of the pump beam of squeezing from the frequency parameter of the system,
$\eta e^{ i \theta}$ ($\eta \geq 0$, $0 \leq \theta < 2\pi$) is the squeezing parameter, 
$\mathcal{D}[L]\rho = L \rho L^{\dag} - (\rho L^{\dag} L - L^{\dag} L \rho)/2$
is the Lindblad form
representing the coupling of the system with the reservoirs through the operator $L$ ($L=a$ or $L=a^{\dag 2}$), 
$\gamma_{1} > 0$ and $\gamma_{2} > 0$ are the decay rates for 
negative damping and nonlinear damping due to coupling 
of the system with the respective reservoirs,
and the reduced Planck constant is set as $\hbar = 1$. 
The two dissipative terms in Eq.(\ref{eq:qvdp_me}) are the quantum analogs of negative damping and nonlinear damping terms in the classical van der Pol model \cite{van1927vii}.

Employing the phase space approach~\cite{gardiner1991quantum, carmichael2007statistical},
we can introduce the Wigner distribution 
$W({\bm \alpha}, t) = \frac{1}{\pi^{2}} \int d \lambda d \lambda^{*} \exp \left(-\lambda \alpha^{*}+\lambda^{*} \alpha\right) \Tr \left\{\rho \exp \left(\lambda a^{\dagger}-\lambda^{*} a\right)\right\}$ 
corresponding to $\rho$ where  
${\bm \alpha} = ( \alpha, \alpha^* )^T \in \mathbb{C}^2$, 
$\lambda,\lambda^* \in \mathbb{C}$, and ${}^*$ indicates complex conjugate.
Then we can derive the following
partial differential equation for $W({\bm \alpha}, t)$:
\begin{align}
  \label{eq:qvdp_fpe}
  \partial_{t} W(\bm{\alpha}, t) =  &- \partial_{\alpha} 
  \left[ \left( \frac{\gamma_1 + 2\gamma_2}{2} + i \Delta \right) \alpha - \gamma_{2} \alpha^{*} \alpha^{2} - 2 \eta e^{i \theta}\alpha^{*} \right]W(\bm{\alpha}, t)
  \cr
  &+ \frac{1}{2} \partial_{\alpha}\partial_{\alpha^{*}} \left[ \frac{\gamma_1}{2} + 2 \gamma_2 \left( \abs{\alpha}^2 -\frac{1}{2} \right) \right]W(\bm{\alpha}, t)
  + \frac{\gamma_2}{4} \partial_{\alpha}^2 \partial_{\alpha^{*}} \alpha W(\bm{\alpha}, t) + H.c.,
\end{align}
where $H.c.$ denotes Hermitian conjugate. 
Note that the third-order derivative terms exist, which are characteristic to quantum systems.
As we discuss below, the nonlinear damping constant $\gamma_2$
controls the intensity of quantum fluctuations in this system.

In the semiclassical regime where $\gamma_1 \gg \gamma_2$, 
the amplitude $\abs{\alpha}$ takes large values on average. 
The third-order derivative terms in Eq.~(\ref{eq:qvdp_fpe}) can then be neglected (see e.g.~\cite{lee2013quantum,lorch2016genuine}) and the coefficients of the second-order derivative terms are positive.
We can thus obtain the following semiclassical SDE in the Ito form: 
\begin{align}
  \label{eq:qvdp_ldv_alp}
  d \left( \begin{matrix}
    \alpha  \\
    \alpha^{*} \\
  \end{matrix} \right)
  = \left(
  \begin{matrix}
    \left(\frac{\gamma_1 + 2\gamma_2}{2} + i \Delta \right) \alpha 
    - \gamma_{2} \alpha^{*} \alpha^{2} - 2 \eta e^{i \theta}\alpha^{*}
    \\
    \left(\frac{\gamma_1 + 2\gamma_2}{2} - i \Delta \right) \alpha^{*}   
    - \gamma_{2} \alpha    \alpha^{*2} - 2 \eta e^{- i \theta}\alpha
    \\
  \end{matrix} 
  \right)dt
  +\sqrt{ \frac{ D(\alpha, \alpha^*) }{2} }
  \begin{pmatrix}
    1 & i \\ 1 & -i 
  \end{pmatrix}
  \begin{pmatrix}
  dw_1 \\ dw_2
  \end{pmatrix},
\end{align}
where $D(\alpha, \alpha^*) = \frac{\gamma_1}{2} + 2 \gamma_2 ( \abs{\alpha}^2 -\frac{1}{2} )$ and 
$w_1$ and $w_2$ are independent Wiener processes satisfying $\mean{dw_{i}(t)dw_{j}(t)} = \delta_{ij} dt$ with $i, j = 1, 2$. 
Representing the complex variable $\alpha$ using the modulus $R$ and argument $\phi$ as $\alpha = R e^{i \phi}$, we obtain the SDEs for these variables as
\begin{align}
  \label{eq:qvdp_ldv_rad}
  dR &= \left( \frac{\gamma_1 + 2\gamma_2}{2} R-\gamma_{2} R^{3} - 2 \eta R \cos (2 \phi-\theta) + Y(R) \right)dt  + \sqrt{ \frac{ D(R) }{2} } dw_R,
  \\
  \label{eq:qvdp_ldv_phi}
  d\phi &= \left( \Delta + 2 \eta \sin (2 \phi-\theta) \right)dt + \frac{1}{R}\sqrt{ \frac{ D(R) }{2} } dw_\phi,
\end{align}
where $w_R$ and $w_{\phi}$ are independent Wiener processes 
satisfying $\mean{dw_{k}(t)dw_{l}(t)} = \delta_{kl} dt$ with $k,l= R,\phi$, 
$D(R) = \frac{\gamma_1}{2} + 2 \gamma_2 ( R^2 -\frac{1}{2} )$, and $Y(R) = \frac{D(R)}{2R}$
is a term arising from the change of the variables by the Ito formula.

Without squeezing, i.e., $\eta  =0$,
the system in the classical limit, described by
the deterministic part of the semiclassical SDE (\ref{eq:qvdp_ldv_alp}),
corresponds to 
a normal form of the supercritical Hopf bifurcation \cite{guckenheimer1983nonlinear}, 
also known as the Stuart-Landau oscillator~\cite{kuramoto1984chemical}
(therefore the quantum van der Pol model is also called the quantum Stuart-Landau model recently \cite{chia2020relaxation,mok2020synchronization,wachtler2020dissipative}).
When the squeezing exists, i.e., $\eta \neq 0$, the system has two stable fixed points for $\Delta \leq 2\eta$ as can be seen from the drift term in Eq.~(\ref{eq:qvdp_ldv_phi}). 
These fixed points annihilate with their unstable counterparts via a SNIC bifurcation at $\Delta = 2 \eta$ and a stable limit-cycle arises when $\Delta > 2\eta$; the argument $\phi$ continuously increases when $\Delta > 2\eta$, while $\phi$ converges to either of the fixed values when $\Delta < 2\eta$.  

When the system in the classical limit is slightly below the SNIC bifurcation, i.e., when $\Delta$ is slightly less than $2 \eta$,
the system can exhibit oscillatory response excited by the noise.
From Eq.~(\ref{eq:qvdp_ldv_rad}), the amplitude of this noisy oscillation is approximately $\mathcal{O}( \sqrt{ \frac{\gamma_1}{\gamma_2}} )$ and therefore $D(R) = \mathcal{O}(\gamma_1)$.
Thus, the intensity of noise acting on $\phi$ is $\mathcal{O}(\frac{1}{R}\sqrt{\frac{ D(R) }{2}}) = \mathcal{O}(\sqrt{\gamma_2})$ and is characterized by the nonlinear damping parameter $\gamma_2$; the larger the value of $\gamma_2$, the stronger the quantum fluctuations acting on the variable $\phi$ are.
In the following analysis, we fix the negative damping parameter $\gamma_1$ and vary $\gamma_2$ to control the intensity of the quantum fluctuations.
We note that fixing $\gamma_1$ to a 
constant value can always be performed by appropriately rescaling the time and other parameters~\cite{kato2019semiclassical}.


\section{Quantum coherence resonance}

\subsection{Semiclassical regime}

First, we numerically analyze the quantum master equation~(\ref{eq:qvdp_me}) in the semiclassical regime  with small nonlinear damping $\gamma_2$. In this regime, we can approximately describe the system by the semiclassical SDEs~(\ref{eq:qvdp_ldv_alp}-\ref{eq:qvdp_ldv_phi}).
Numerical simulations are performed by using QuTiP numerical toolbox~\cite{johansson2012qutip,*johansson2013qutip}.
We define the autocovariance and normalized power spectrum of the system as
\begin{align}
\label{eq:nspe}
  C(\tau) = \mean{ a^{\dag}( \tau ) a(0)} - \mean{a^{\dag}( \tau )} \mean{a(0)},
  \quad
  \bar{S}(\omega) &= \int_{-\infty}^{\infty} d\tau e^{i\omega \tau} 
  C(\tau)/C(0),
\end{align}
where $\mean{A} = \Tr{[ A\rho]}$ is the expectation value of $A$
with respect to $\rho$ in the steady state.
We set the parameters such that
the system in the classical limit is near a SNIC bifurcation and vary $\gamma_2$.
We fix the parameter $\gamma_1$ at $\gamma_{1} = 1$ without loss of generality, as it can be eliminated by rescaling the time in the master equation~(\ref{eq:qvdp_me}).

In classical coherence resonance, the regularity
of the system's response is quantitatively characterized 
from the (non-normalized) power spectrum by 
$\beta =  h \omega_{p}/ (\Delta \omega_{h})$,
where $\omega_{p}$, $h$, and $\Delta \omega_{h}$ 
are the peak frequency, peak height,  
and width of the spectrum, respectively~\cite{gang1993stochastic}. 
In the present case, the amplitude of the response varies with the parameter $\gamma_2$, because
the deterministic part of the SDE~(\ref{eq:qvdp_ldv_alp}) 
explicitly depends on $\gamma_2$. 
Since the amplitude of the response is not relevant to the regularity of the response, we define a degree of coherence by using the normalized power spectrum as
\begin{align}
\label{eq:deg_coh}
\bar{\beta} &=  \bar{S}(\omega_{p}) \omega_{p}/ (\Delta \omega),
\end{align}
where $\omega_{p} = \argmax_{\omega}\bar{S}(\omega)$, $\bar{S}(\omega_{p})$, and
$\Delta \omega$ are the peak frequency, peak height, and full width at half
maximum of $\bar{S}(\omega)$, respectively, and we use this $\bar{\beta}$ to quantify the regularity of the response.

\begin{figure} [!t]
	\begin{center}
		\includegraphics[width=1\hsize,clip]{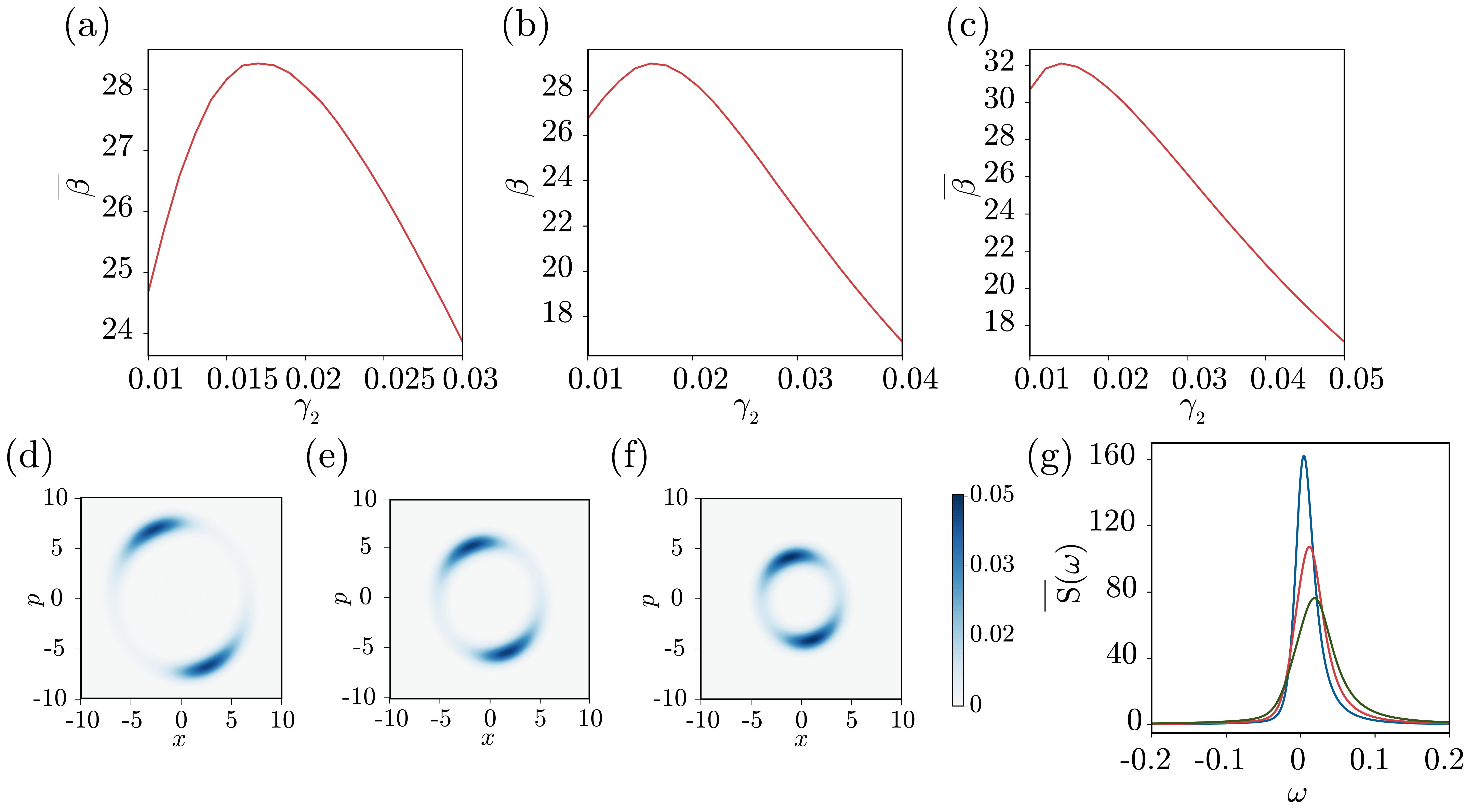}
		\caption{
			Results in the semiclassical regime.
			(a-c) 
			Normalized degree of coherence $\bar{\beta}$ vs. nonlinear damping constant $\gamma_2$ (intensity of the quantum fluctuations). 
			(d-f):
			Steady-state Wigner distributions for $\gamma_2 = 0.01$(d), $0.017$(e), $0.03$(f).
			(g): Power spectra for $\gamma_2 = 0.01$ (blue), 
			$0.017$ (red), and
			$0.03$ (green).
			The other parameters are 
			$\eta e^{i \theta} = 0.025$, $\Delta = 0.0375$ (a,d-g),	
			$\eta e^{i \theta} = 0.02$, $\Delta = 0.0275$ (b),
			and 
			$\eta e^{i \theta} = 0.03$, $\Delta = 0.05$
			 (c).
			The negative damping rate is fixed at $\gamma_1 = 1$.
		}
		\label{fig_1}
	\end{center}
\end{figure}

\begin{figure} [!t]
	\begin{center}
		\includegraphics[width=1\hsize,clip]{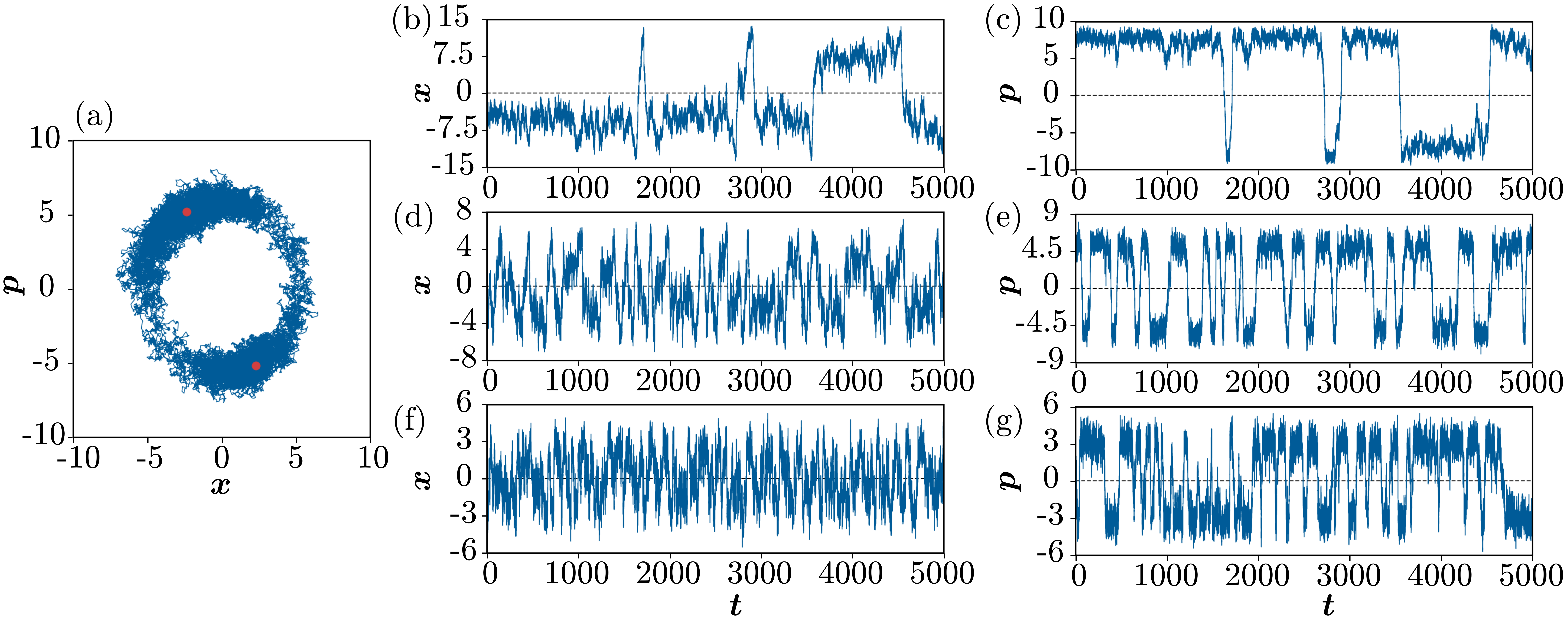}
		\caption{
			Time evolution of a single trajectory of the semiclassical SDE~(\ref{eq:qvdp_ldv_alp}) 
			with 
			$\eta e^{i \theta} = 0.025$, $\Delta = 0.0375$ and
			$\gamma_1 = 1$.
			(a): Trajectory on the $x-p$ plane ($x= \mbox{Re}~\alpha$ and $p = \mbox{Im}~\alpha$)
			at $\gamma_2 = 0.017$ for $0 \leq t \leq 1000$~(blue curve).
			Stable fixed points in the classical limit are indicated by the red dots.
			(b,d,f): Time evolution of $x$ for
			$\gamma_2 = 0.003$ (b), $0.017$ (d), and $0.05$ (f).
			(c,e,g): Time evolution of $p$ for
			$\gamma_2 = 0.003$ (c), $0.017$ (e), and $0.05$ (g).
		}
		\label{fig_2}
	\end{center}
\end{figure}

Figures~\ref{fig_1}(a)-\ref{fig_1}(c) show the dependence of the degree of coherence $\bar{\beta}$ on the nonlinear damping constant $\gamma_2$, i.e., on the intensity of quantum fluctuations in the semiclassical regime for three different parameter settings.
We can observe that $\overline{\beta}$ takes a maximum value at a certain value of $\gamma_2$ in each figure. The locations of these peaks are $\gamma_2 = 0.017, 0.016$ and $0.014$ in Fig.~\ref{fig_1}(a), ~\ref{fig_1}(b), and (c), respectively. 
This result indicates that there exists an optimal intensity of the quantum fluctuations that maximizes the regularity of the oscillatory response,
namely, the quantum coherence resonance occurs in the present system in this semiclassical regime.

The steady-state Wigner distributions at $\gamma_2 = 0.01$, $0.017$, and $0.03$ 
for the parameter setting used in Fig~\ref{fig_1}(a)
are shown in Figs.~\ref{fig_1}(d)-~\ref{fig_1}(f). In each figure, the distribution is localized around the two stable fixed points on an ellipse connecting them in the classical limit.
The normalized power spectra $\bar{S}(\omega)$ of the system are shown in Fig.~\ref{fig_1}(g) for three values of $\gamma_2$.
Note that the amplitude of the noise-induced oscillations depends on $\gamma_{2}$; the normalization of the power spectrum reduces this dependence and allows us to focus on the regularity of the oscillations.
The dependence of the power spectrum on $\gamma_2$ in Fig.~\ref{fig_1}(g)
is similar to that for coherence resonance of a classical FitzHugh-Nagumo system 
in the bistable excitable regime discussed in~\cite{lindner2000coherence, *lindner2002coherence},
which has a peak at a positive value of $\omega$ reflecting that counter-clockwise stochastic rotations of the system trajectory in the phase-space representation are dominant in this regime. 

Time evolution of a single trajectory of 
the semiclassical SDE~(\ref{eq:qvdp_ldv_alp})
after the initial transient is shown in Fig.~\ref{fig_2}.
Figure~\ref{fig_2}(a) shows the trajectory at $\gamma_2  = 0.017$ on the $x-p$ plane with $x = \mbox{Re}\ \alpha$ and $p = \mbox{Im}\ \alpha$, corresponding to the steady-state Wigner distribution in Fig.~\ref{fig_1}(e) for which the semiclassical approximation is valid.
As the system in the classical limit is slightly below the SNIC bifurcation,
it has two stable fixed points represented by the red dots.
The quantum fluctuations induce stochastic oscillations between these two points
by kicking the system state out of these fixed points, as shown by the
blue curve.

On the analogy of classical coherence resonance in noisy excitable systems, the quantum coherence resonance phenomenon in the semiclassical regime observed 
above can be understood as follows.
Figures~\ref{fig_2}(b,d,f) and (c,e,g) show the time evolution of 
$x$ and $p$ for the three cases with $\gamma_2 = 0.003$ (b, c), 
$0.017$ (d, e), and $0.05$ (f, g), respectively.
When $\gamma_2 = 0.003$, the quantum noise is too weak and the system state hardly takes a round trip around the two stable fixed points. The response of the system is weak
and irregular.
For intermediate noise intensity with $\gamma_2 = 0.017$,
the noise excites round trips of the system state more frequently,
leading to the more regular response.
When $\gamma_2 = 0.05$,
the noise is too strong and induces irregularity of the response.
These results provide a semiclassical interpretation of the existence of the optimal value of $\gamma_2$ in Fig.~\ref{fig_1}(a), which was obtained from the quantum master equation in Eq.~(\ref{eq:qvdp_me}).

It is interesting to note that the round trips in Fig.~\ref{fig_2} are induced by the noise representing quantum fluctuations. Therefore, quantum coherence resonance in this regime can also be interpreted as a noise-enhanced quantum tunneling, which is similar to the quantum tunneling effects observed in the quantum stochastic resonance \cite{grifoni1996coherent, grifoni1996quantum} 
and in the transition between metastable states of the dispersive 
optical bistability \cite{risken1987quantum,risken1988quantum,vogel1988quantum}.

\subsection{Weak quantum regime}

\begin{figure} [!t]
  \begin{center}
    \includegraphics[width=1\hsize,clip]{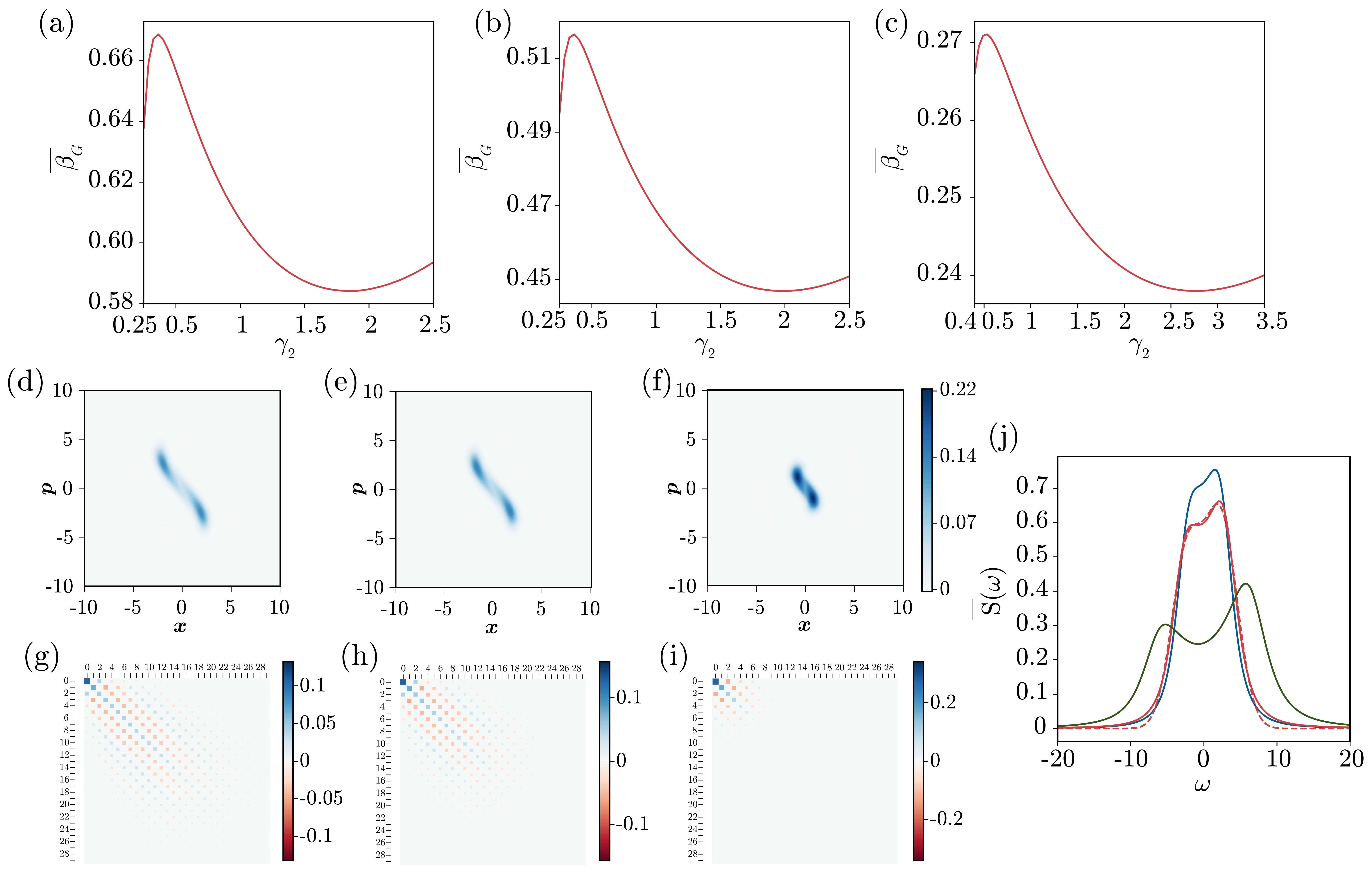}
    \caption{
    	Results in the weak quantum regime.
    	(a-c): 
    	Normalized degree of coherence $\bar{\beta}_G$ vs. nonlinear damping constant $\gamma_2$ (intensity of the quantum fluctuations).
    	(d-f):
    	Wigner distributions for $\gamma_2 = 0.4$ (d), $0.53$ (e), $2.75$ (f).
    	(g-i): 
		Elements of the density matrix with respect to the number basis
    	for $\gamma_2 = 0.4$ (g), $0.53$ (h), $2.75$ (i).
    	(j): Power spectra for $\gamma_2 = 0.4$ (thin blue), 
    	$\gamma_2 =0.53$ (thin red) with the approximated Gaussian mixed  model (dotted red), and $\gamma_2 = 2.75$ (thin green).
		The other parameters are $\eta e^{i \theta} = 2$,
	  	$\Delta = 3.825$
		 (a),
	  	$\eta e^{i \theta} = 3$, $\Delta = 5.8$
		(b),
		and
		$\eta e^{i \theta} = 7$, $\Delta = 13.65$
		(c)-(m). 
	The parameter $\gamma_1$ is fixed at $1$. 
}
    \label{fig_3}
  \end{center}
\end{figure}

Next, we consider the weak quantum regime with moderately strong quantum fluctuations, where the nonlinear damping $\gamma_2$ and the detuning frequency $\Delta$ are larger than the previous semiclassical regime.
In this regime, the power spectrum generally takes two distinct peaks 
at $\omega_p$ and $-\omega_p$ (see Fig.~\ref{fig_3}(j)) and we cannot simply measure the total width of these two overlapped peaks since it may yield inappropriate values of the normalized degree of coherence $\overline{\beta}$. Therefore, we evaluate the value of $\overline{\beta}$ using only the highest peak of the power spectrum.
To this end, we fit the normalized power spectrum by the Gaussian mixture model 
\begin{align}
\label{gauss1}
\bar{S}(\omega) \approx h_1 \exp \left\{-\frac{(\omega - \bar{\omega})^{2}}{2 \sigma_1^{2}}\right\}
+ h_2 \exp \left\{-\frac{(\omega + \bar{\omega})^{2}}{2 \sigma_2^{2}}\right\}
\end{align}
where $h_1$ and $h_2$ ($h_1 \geq h_2$) are the heights, $\sigma_{1,2}$ are the standard deviations, and $\pm \bar{\omega}$ are the mean values of the Gaussian distributions,
and approximately evaluate the normalized degree of coherence as 
\begin{align}
\label{eq:deg_coh_gauss}
\bar{\beta} \approx \bar{\beta}_G = h_1 \bar{\omega} / \Delta \bar{\omega}
\end{align}
with $\Delta \bar{\omega} = 2 \sigma_1 \sqrt{2 \ln 2}$ 
representing the full width at half maximum of the Gaussian distribution with standard deviation $\sigma_1$. 
Using this quantity, we can appropriately measure the normalized degree of coherence even when the normalized power spectrum has two peaks.

Dependence of 
$\bar{\beta}_G$ on
$\gamma_2$ are shown in Figs.~\ref{fig_3}(a-c) for three different parameter settings.
It is remarkable that $\bar{\beta}_G$ also has a peak at a certain value of $\gamma_2$ in all figures.
The peaks occur at $\gamma_2 = 0.36$, $0.36$, and $0.53$, respectively.	
Thus, we observe quantum coherence resonance also in this quantum regime with moderate quantum fluctuations.
Here, we stress that the semiclassical SDEs are no longer valid at these values of $\gamma_2$.
As $\gamma_2$ becomes larger, truncation of the third-order derivative terms in Eq.~(\ref{eq:qvdp_fpe}) and approximation by the SDE~(\ref{eq:qvdp_ldv_alp}) become less accurate, and further increase in $\gamma_2$ leads to negative values of $D$, for which the approximation by the SDE~(\ref{eq:qvdp_ldv_alp}) is no longer possible.
The second increase in $\bar{\beta}_G$ at large $\gamma_2$ in Figs.~\ref{fig_3}(a)-~\ref{fig_3}(c) is due to strong quantum effects and will be discussed in the next subsection.

Wigner distributions, 
elements of the density matrix $\rho$ with respect to the number basis, and normalized power spectra  
in the steady state 
are shown in Figs.~\ref{fig_3}(d)-\ref{fig_3}(f),
Figs.~\ref{fig_3}(g)-\ref{fig_3}(i), and Fig.~\ref{fig_3}(j), respectively,
for the parameter setting used in Fig.~\ref{fig_3}(c)
with $\gamma_2 = 0.4$, $0.53$, and $2.75$.
Note that $\gamma_2 = 0.53$ and $2.75$ give the maximum and minimum of $\bar{\beta}_G$ in Fig.~\ref{fig_3}(c), respectively. 

The Wigner distributions are localized around the two fixed points as in the semiclassical case, 
but the ellipse connecting them is strongly compressed and deformed, 
reflecting the quantum effect;
they are concentrated in the phase-space region where the average 
number of photons are smaller. 
However, as can be seen from Figs.~\ref{fig_3}(g)-~\ref{fig_3}(i), the elements of $\rho$ still take non-zero values 
up to considerably high energy levels, suggesting that the discreteness of the energy spectrum is still not dominant in this regime.

The power spectra shown in Fig.~\ref{fig_3}(j) differ from those in the semiclassical regime (Fig.~\ref{fig_1}(g)) in several aspects. 
First, the peak heights of the power spectra in this regime are two orders of magnitude smaller than the previous semiclassical regime, because the system is in the lower energy states with stronger quantum fluctuations on average.
Second, as we stated previously, the power spectra in this regime have two distinct peaks, in contrast to those with a single peak in the semiclassical regime.
Despite these differences, the overall dependence of the normalized power spectrum on $\gamma_2 $ in Fig.~\ref{fig_3}(j) is qualitatively 
similar to those for the semiclassical case shown in Fig.~\ref{fig_1}(g).

\begin{figure} [!t]
	\begin{center}
		\includegraphics[width=0.75\hsize,clip]{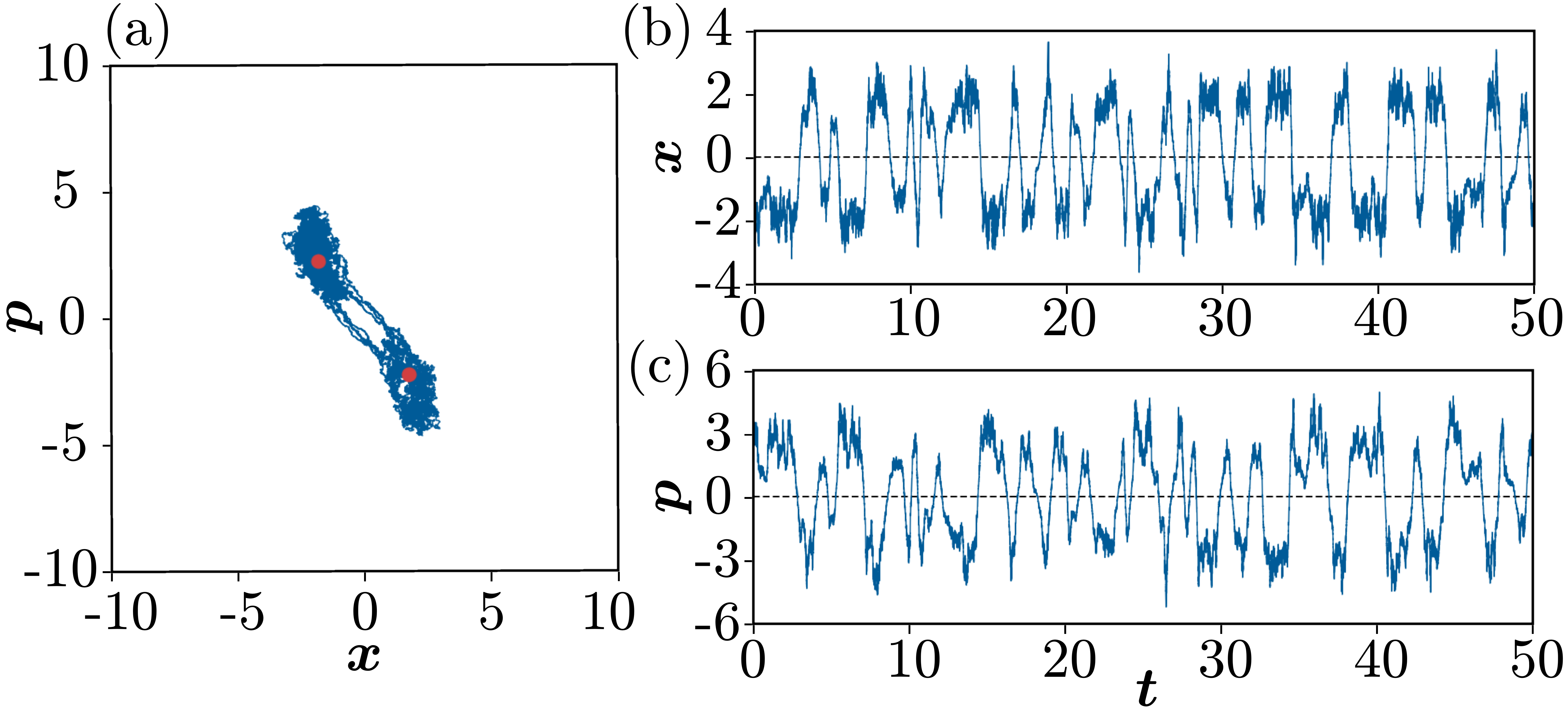}
		\caption{
			Time evolution of a single trajectory of the approximate semiclassical SDE~(\ref{eq:qvdp_ldv_alp}) 
			in the weak quantum regime with 
			$\eta e^{i \theta} = 7$, $\Delta = 13.65$, 
			$\gamma_1 = 1$, and $\gamma_2 = 0.5$.
			(a): 
			Trajectory on the $x-p$ plane for $0 \leq t \leq 10$~(blue curve).
			Stable fixed points in the classical limit are indicated by red dots.
			(b): Time evolution of $x$.
			(c): Time evolution of $p$.
		}
		\label{fig_4}
	\end{center}
\end{figure}

Although the approximation by the semiclassical SDE~(\ref{eq:qvdp_ldv_alp}),
which is derived by neglecting the third-order derivatives in Eq.~(\ref{eq:qvdp_fpe}),
is quantitatively inaccurate in the present case,  
we can still depict the time evolution of a single trajectory of the approximate SDE
as long as $D$ takes a positive value, which helps us obtain a qualitative picture
of the system dynamics in this regime.
This is possible up to $\gamma_2 \approx 0.5$ where $D$ remains positive in the present system; 
at $\gamma_2=0.53$ where $\bar{\beta}_G$ takes the peak value, $D$ becomes negative and we can no longer consider the classical trajectory even in the approximate sense.

Figure~\ref{fig_4} shows
approximate evolution of a single trajectory obtained as above 
for $\gamma_2 = 0.5$ after the initial transient.
In Fig.~\ref{fig_4}(a), the trajectory is plotted on the $x-p$ plane.
As the system in the classical limit is slightly below the SNIC bifurcation, 
there are two stable fixed points represented by the red dots,
and the system exhibits stochastic oscillations between these two points
as shown by the blue curve.
Note that the dynamics is much faster than the previous semiclassical case in Fig.~\ref{fig_2}.
The trajectory is strongly deformed and appears to be jumping between the two fixed points, in contrast to the semiclassical case where it was ellipsoidal.
The time evolution of $x$ and $p$ in this case are plotted in Figs.~\ref{fig_4}(b,c),
showing noisy switching between the two fixed points induced by the quantum fluctuations.

The appearance of double peaks in the power spectrum (Fig.~\ref{fig_3}(j)) can be understood as follows.
In the previous semiclassical case, the system exhibited circular, counter-clockwise stochastic rotations, and this directionality of rotation is reflected in the power spectrum containing mainly positive-frequency components peaked at the characteristic frequency $\omega_p$ (Figs.~\ref{fig_1} and~\ref{fig_2}).
In the present case, in contrast, the effect of squeezing leads to strongly asymmetric Wigner distributions and nearly straight-line approximate stochastic trajectories connecting the two fixed points in the classical limit as observed from Figs.~\ref{fig_3}(d-f) and~\ref{fig_4}.
Consequently, the power spectrum contains comparably large
negative frequency components around $- \omega_p$ in addition to the
positive-frequency components around $\omega_p$.
A small additional peak at a negative frequency $-\omega_p$ in the power spectrum on top of a large peak at $\omega_p$ is also observed in the asymmetric limit-cycle oscillations of the quantum van der Pol oscillator subjected to squeezing in the semiclassical regime \cite{kato2019semiclassical}.

It should be stressed that, though the dynamics of the system in these quantum regimes with relatively large $\gamma_2$ cannot be accurately described by the semiclassical picture, the degree of coherence $\bar{\beta}_G$ obtained by direct numerical simulations of the master equation clearly takes a peak value when plotted against $\gamma_2$ as shown in Fig.~\ref{fig_3}(a)-(c).
These results indicate that the present system exhibits coherence resonance also in the quantum regime with moderately strong quantum fluctuations, where no semiclassical counterpart exists.

\begin{figure} [!t]
	\begin{center}
		\includegraphics[width=0.95\hsize,clip]{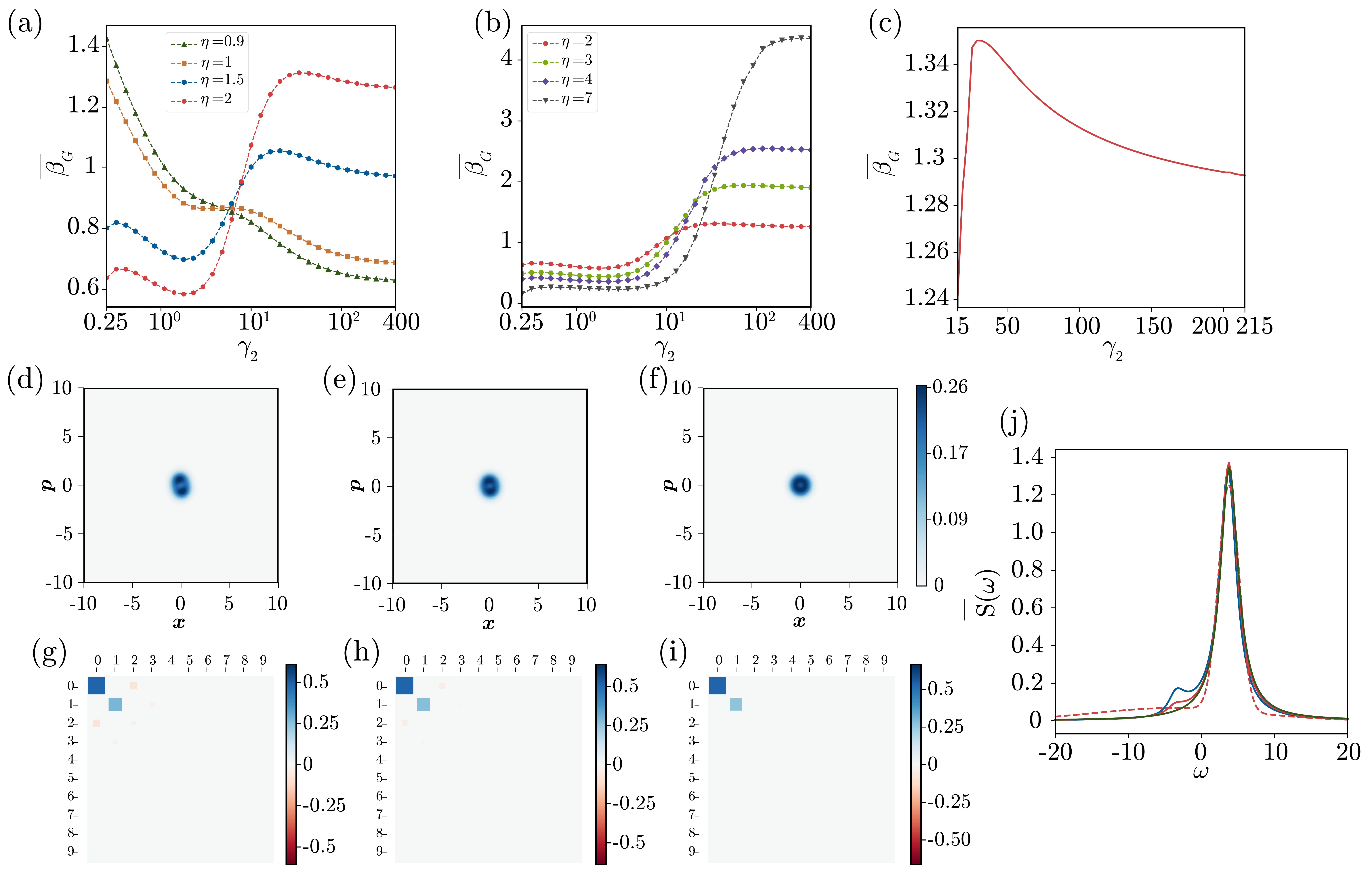}
		\caption{
			Results in the strong quantum regime. 
			(a)-(c): 
			Normalized degree of coherence $\bar{\beta}_G$ 
			vs. nonlinear damping constant ( intensity of the quantum fluctuations) $\gamma_2$.
			(d-f):
			Wigner distributions for $\gamma_2 = 15$ (d), $28.3$ (e), $215$ (f).
			(g-i): Elements of the density matrix with respect to the number basis
			for $\gamma_2 = 15$ (g), $28.3$ (h), $215$ (i).
			(j): Power spectra for $\gamma_2 = 15$ (blue), 
			$28.3$ (red) with the approximated Gaussian mixed model (red-dot), and $215$ (green).	
			In (a),
			the other parameters are 
			$\eta  = 0.9$, $\Delta = 1.75$
			(dark green),
			$\eta  = 1$, $\Delta = 1.93$
			(brown),
			$\eta  = 1.5$, $\Delta = 2.85$
			(blue),
			$\eta  = 2$, $\Delta = 3.825$
			(red).
			In (b),
			the other parameters are
			$\eta  = 2$, $\Delta = 3.825$
			(red),
			$\eta  = 3$, $\Delta = 5.8$
			(light green),
			$\eta  = 4$, $\Delta = 7.775$
			(purple),
			$\eta  = 7$, $\Delta = 13.65$
			(dark gray).
			In (c)-(j), the parameters are  
			$\eta = 2$ and $\Delta = 3.825$.
			The parameter $\gamma_1$ is fixed at $1$ and $\theta$ is fixed at $0$.}
		\label{fig_5}
	\end{center}
\end{figure}

\subsection{Strong quantum regime}

We now consider the strong quantum regime with much larger nonlinear damping $\gamma_2$, where only a small number of energy states participate in the system dynamics.

First, Figs.~\ref{fig_5}(a) and \ref{fig_5}(b) show the overall dependence of the degree of coherence $\bar{\beta}_G$ on $\gamma_2$ for several parameter settings of the 
squeezing $\eta e^{i\theta}$ and detuning $\Delta$  chosen appropriately, including those used in Fig.~\ref{fig_3}. 
Note that the range of $\gamma_2$ is much wider than those in Fig.~\ref{fig_3} and the dependence on $\gamma_2$ is plotted on the logarithmic scale.
It is remarkable that $\bar{\beta}_G$ exhibits the second peak around $\gamma_2 = 28.3$ when the squeezing parameter is $\eta = 2$. 
As we decrease $\eta$, this second peak becomes weaker and finally disappears when $\eta = 0.9$ as can be seen in Fig.~\ref{fig_5}(a).
When we increase $\eta$, this second peak tends to be flattened,  
but it still persists at $\eta = 7$ as can be seen in Fig.~\ref{fig_5}(b). 
Figure~\ref{fig_5}(c) shows the enlargement of the region near the second peak of $\bar{\beta}_G$ for $\eta=2$ on a linear scale.

It should be stressed that this second peak cannot be interpreted using the semiclassical picture because, if the semiclassical picture is valid, increasing the intensity of quantum fluctuations beyond the first peak of the coherence resonance simply destroys the regularity of the system.
Therefore, we should regard this second peak as an explicit quantum effect arising from the negative diffusion constant and the third-order derivative term in Eq.~(\ref{eq:qvdp_fpe}).

The Wigner distributions, elements of the density matrix, and normalized power spectra in the steady state are shown in Figs.~\ref{fig_5}(d)-~\ref{fig_5}(f),
Figs.~\ref{fig_5}(g)-~\ref{fig_5}(i), and Fig.~\ref{fig_5}(j), respectively,
where the intensities of quantum fluctuations (values of the nonlinear damping constant) are $\gamma_2 = 15$, $28.3$, and $215$, and the other parameters are the same as in the case of $\eta = 2$ shown in Fig.~\ref{fig_3}(a).

The Wigner distributions are localized around the two classical fixed points as in the previous cases, which are now very close to the origin because 
the system is in the lower energy state 
as a result of the strong nonlinear damping.
The Wigner distributions is slightly asymmetric at $\gamma_2=15$ (Fig.~\ref{fig_5}(d)) and tends to be symmetric as $\gamma_2$ is further increased (Figs.~\ref{fig_5}(e,f)) and the classical fixed points approach each other.
We can also observe in Figs.~\ref{fig_5}(g)-~\ref{fig_5}(i) that the matrix elements of $\rho$ are mostly close to zero except for several elements whose energy levels are close to the ground state.
This indicates that only a few energy levels participate in the system dynamics and the discreteness of the energy spectra 
can play dominant roles in this regime.

The steady-state density matrix possesses several non-small elements at $\gamma_2 = 15$ as shown in Fig.~\ref{fig_5}(g), which corresponds to the Wigner distribution in Fig.~\ref{fig_5}(d).
At $\gamma_2 = 28.3$ around the second peak,
only four matrix elements, namely, those at $\ket{0}\bra{0}$, $\ket{1}\bra{1}$, $\ket{2}\bra{0}$, and $\ket{0}\bra{2}$ representing transitions among the lowest three energy states $\ket{0}, \ket{1}$, and $\ket{2}$ become dominant as shown in Figs.~\ref{fig_5}(h).
When $\gamma_2 = 215$ (Fig.~\ref{fig_5}(i)), only the matrix elements at $\ket{0}\bra{0}$ and $\ket{1}\bra{1}$ survive and all other matrix elements are close to zero.
Indeed, 
the steady-state density matrix practically approaches that of the quantum vdP system in the strong quantum limit without squeezing, $\rho \approx 2/3\ket{0}\bra{0} + 1/3\ket{1}\bra{1}$, 
in the limit of large nonlinear damping $\gamma_2$~\cite{lee2013quantum},
because the system state approaches the ground state with the increase of the nonlinear damping $\gamma_2$. 
This result suggests that the transitions between the ground state, the single-photon state, and the two-photon state can exhibit strong resonance at the appropriate intensity of the quantum fluctuations when the effect of the squeezing is strong, yielding the second peak in the degree of coherence.

The dependence of the normalized power spectrum on $\gamma_2$ in Fig.~\ref{fig_5}(j) is qualitatively similar to those of the previous cases in Fig.~\ref{fig_1}(g) and Fig.~\ref{fig_3}(j).
In this strongly quantum regime, the peak at $-\omega_p$ tends to be lower than the peak at $\omega_p$ and becomes almost invisible at $\gamma_2 = 215$.

From the results for the weak quantum regime in Fig.~\ref{fig_3} and for the strong quantum regime in Fig.~\ref{fig_5}, we conclude that we can observe two peaks in the degree of coherence in the quantum regime with appropriate parameter settings as $\gamma_2$ is increased, where the first peak corresponds to the coherence resonance also observed in the semiclassical regime, while the second peak is caused by the strong quantum effect.


\section{Discussion}

We have demonstrated quantum coherence resonance
in a quantum van der Pol system subjected to squeezing.
In the semiclassical regime, we could interpret this phenomenon
on the analogy of classical noisy excitable systems 
using SDEs describing the phase-space trajectory.
We also confirmed that this phenomenon persists under moderately strong quantum fluctuations for which the semiclassical description is not valid but still a large number of energy levels contribute to the system dynamics.
Moreover, we demonstrated that the system can exhibit the second peak in the degree of coherence as the intensity of quantum fluctuations is further increased, where only a small number of energy levels participate in the dynamics and strong quantum effect dominates the system.

We think it possible to observe quantum coherence resonance, in principle, in the currently available experimental setup. 
The quantum van der Pol model subjected to squeezing can be experimentally implemented in the ``membrane-in-the-middle'' optomechanical setup, where the negative damping and nonlinear damping terms 
can be realized by
applying lasers detuned to the blue one-photon sideband and red two-photon sideband, respectively \cite{walter2014quantum, sonar2018squeezing}, and the squeezing term can be realized by electrically modulating the spring constant at twice the mechanical frequency \cite{rugar1991mechanical,sonar2018squeezing}. 
Alternatively, it may also be possible to use ion-trap systems for the
experimental realization of the model in the strong quantum regime \cite{lee2013quantum}. 
The quantum coherence resonance can be observed by varying the nonlinear damping parameter and measuring the power spectrum
to evaluate the normalized degree of coherence in these experimental setups. 
	
In quantum coherence resonance, the regularity of system's response
is enhanced by the constructive effect of the quantum fluctuations.
This is in contrast to the case of quantum synchronization discussed
in the previous studies, where the quantum fluctuations had deleterious effect
on the quality of synchronization~\cite{lee2013quantum, walter2014quantum, sonar2018squeezing, kato2019semiclassical}.
The relation between the regularity of the system's oscillatory response 
and the intensity of quantum fluctuations determined by the coupling constants with the reservoirs
would provide a guideline for designing experimental
setups that realize quantum coherence resonance.
As a further generalization, it would also be interesting to investigate quantum
coherence resonance in 
networks of quantum excitable systems by
extending past studies on networks of classical excitable systems~\cite{han1999interacting, neiman1999noise, toral2003system, andreev2018coherence}. 

The quantum coherence resonance could bring new insights into
possible future applications of quantum dissipative systems
in the growing fields of quantum technologies, such as quantum information, quantum metrology, and quantum standard.

We gratefully acknowledge stimulating discussions with N. Yamamoto. This research was financially supported by the JSPS KAKENHI Grant Numbers JP17H03279, JPJSBP120202201, JP20J13778 and JP18H03287, and JST CREST JP-MJCR1913.


\begin{thebibliography}{54}%
\makeatletter
\providecommand \@ifxundefined [1]{%
\@ifx{#1\undefined}
}%
\providecommand \@ifnum [1]{%
\ifnum #1\expandafter \@firstoftwo
\else \expandafter \@secondoftwo
\fi
}%
\providecommand \@ifx [1]{%
\ifx #1\expandafter \@firstoftwo
\else \expandafter \@secondoftwo
\fi
}%
\providecommand \natexlab [1]{#1}%
\providecommand \enquote  [1]{``#1''}%
\providecommand \bibnamefont  [1]{#1}%
\providecommand \bibfnamefont [1]{#1}%
\providecommand \citenamefont [1]{#1}%
\providecommand \href@noop [0]{\@secondoftwo}%
\providecommand \href [0]{\begingroup \@sanitize@url \@href}%
\providecommand \@href[1]{\@@startlink{#1}\@@href}%
\providecommand \@@href[1]{\endgroup#1\@@endlink}%
\providecommand \@sanitize@url [0]{\catcode `\\12\catcode `\$12\catcode
`\&12\catcode `\#12\catcode `\^12\catcode `\_12\catcode `\%12\relax}%
\providecommand \@@startlink[1]{}%
\providecommand \@@endlink[0]{}%
\providecommand \url  [0]{\begingroup\@sanitize@url \@url }%
\providecommand \@url [1]{\endgroup\@href {#1}{\urlprefix }}%
\providecommand \urlprefix  [0]{URL }%
\providecommand \Eprint [0]{\href }%
\providecommand \doibase [0]{http://dx.doi.org/}%
\providecommand \selectlanguage [0]{\@gobble}%
\providecommand \bibinfo  [0]{\@secondoftwo}%
\providecommand \bibfield  [0]{\@secondoftwo}%
\providecommand \translation [1]{[#1]}%
\providecommand \BibitemOpen [0]{}%
\providecommand \bibitemStop [0]{}%
\providecommand \bibitemNoStop [0]{.\EOS\space}%
\providecommand \EOS [0]{\spacefactor3000\relax}%
\providecommand \BibitemShut  [1]{\csname bibitem#1\endcsname}%
\let\auto@bib@innerbib\@empty
\bibitem [{\citenamefont {Horsthemke}(1984)}]{horsthemke1984noise}%
\BibitemOpen
\bibfield  {author} {\bibinfo {author} {\bibfnamefont {W.}~\bibnamefont
{Horsthemke}}\ and\ \bibinfo {author} {\bibfnamefont {R.}~\bibnamefont
{Lefever}},}\bibfield  {title} {\emph {\bibinfo {title} {
{Noise
induced transitions}}}\ }(\bibinfo  {publisher} {Springer},\ \bibinfo
{year} {1984})\BibitemShut {NoStop}%
%
\bibitem [{\citenamefont {Schimansky-Geier}\ \emph {et~al.}(1998)\citenamefont
{Schimansky-Geier}, \citenamefont {Freund}, \citenamefont {Neiman},\ and\
\citenamefont {Shulgin}}]{schimansky1998noise}%
\BibitemOpen
\bibfield  {author} {\bibinfo {author} {\bibfnamefont {L.}~\bibnamefont
{Schimansky-Geier}}, \bibinfo {author} {\bibfnamefont {J.~A.}\ \bibnamefont
{Freund}}, \bibinfo {author} {\bibfnamefont {A.~B.}\ \bibnamefont {Neiman}},
\ and\ \bibinfo {author} {\bibfnamefont {B.}~\bibnamefont {Shulgin}},\
}\bibfield  {title} {\enquote {\bibinfo {title} {Noise induced order:
Stochastic resonance},}\ }\href@noop {} {\bibfield  {journal} {\bibinfo
{journal} {International Journal of Bifurcation and Chaos}\ }\textbf
{\bibinfo {volume} {8}},\ \bibinfo {pages} {869--879} (\bibinfo {year}
{1998})}\BibitemShut {NoStop}%
\bibitem [{\citenamefont {Lindner}\ \emph {et~al.}(2004)\citenamefont
{Lindner}, \citenamefont {Garc{\i}a-Ojalvo}, \citenamefont {Neiman},\ and\
\citenamefont {Schimansky-Geier}}]{lindner2004effects}%
\BibitemOpen
\bibfield  {author} {\bibinfo {author} {\bibfnamefont {B.}~\bibnamefont
{Lindner}}, \bibinfo {author} {\bibfnamefont {J.}~\bibnamefont
{Garc{\i}a-Ojalvo}}, \bibinfo {author} {\bibfnamefont {A.}~\bibnamefont
{Neiman}}, \ and\ \bibinfo {author} {\bibfnamefont {L.}~\bibnamefont
{Schimansky-Geier}},\ }\bibfield  {title} {\enquote {\bibinfo {title}
{Effects of noise in excitable systems},}\ }\href@noop {} {\bibfield
{journal} {\bibinfo  {journal} {Physics Reports}\ }\textbf {\bibinfo {volume}
{392}},\ \bibinfo {pages} {321--424} (\bibinfo {year} {2004})}\BibitemShut
{NoStop}%
\bibitem [{\citenamefont {Goldobin}\ and\ \citenamefont
{Pikovsky}(2005)}]{goldobin2005synchronization}%
\BibitemOpen
\bibfield  {author} {\bibinfo {author} {\bibfnamefont {D.~S.}\ \bibnamefont
{Goldobin}}\ and\ \bibinfo {author} {\bibfnamefont {A.~S.}\ \bibnamefont
{Pikovsky}},\ }\bibfield  {title} {\enquote {\bibinfo {title}
{Synchronization of self-sustained oscillators by common white noise},}\
}\href@noop {} {\bibfield  {journal} {\bibinfo  {journal} {Physica A:
Statistical Mechanics and its Applications}\ }\textbf {\bibinfo {volume}
{351}},\ \bibinfo {pages} {126--132} (\bibinfo {year} {2005})}\BibitemShut
{NoStop}%
\bibitem [{\citenamefont {Teramae}\ and\ \citenamefont
{Tanaka}(2004)}]{teramae2004robustness}%
\BibitemOpen
\bibfield  {author} {\bibinfo {author} {\bibfnamefont {J.~N.}\ \bibnamefont
{Teramae}}\ and\ \bibinfo {author} {\bibfnamefont {D.}~\bibnamefont
{Tanaka}},\ }\bibfield  {title} {\enquote {\bibinfo {title} {Robustness of
the noise-induced phase synchronization in a general class of limit cycle
oscillators},}\ }\href@noop {} {\bibfield  {journal} {\bibinfo  {journal}
{Physical Review Letters}\ }\textbf {\bibinfo {volume} {93}},\ \bibinfo
{pages} {204103} (\bibinfo {year} {2004})}\BibitemShut {NoStop}%
\bibitem [{\citenamefont {Nakao}\ \emph {et~al.}(2007)\citenamefont {Nakao},
\citenamefont {Arai},\ and\ \citenamefont {Kawamura}}]{nakao2007noise}%
\BibitemOpen
\bibfield  {author} {\bibinfo {author} {\bibfnamefont {H.}~\bibnamefont
{Nakao}}, \bibinfo {author} {\bibfnamefont {K.}~\bibnamefont {Arai}}, \ and\
\bibinfo {author} {\bibfnamefont {Y.}~\bibnamefont {Kawamura}},\ }\bibfield
{title} {\enquote {\bibinfo {title} {Noise-induced synchronization and
clustering in ensembles of uncoupled limit-cycle oscillators},}\ }\href@noop
{} {\bibfield  {journal} {\bibinfo  {journal} {Physical Review Letters}\
}\textbf {\bibinfo {volume} {98}},\ \bibinfo {pages} {184101} (\bibinfo
{year} {2007})}\BibitemShut {NoStop}%
\bibitem [{\citenamefont {Toral}\ \emph {et~al.}(2001)\citenamefont {Toral},
\citenamefont {Mirasso}, \citenamefont {Hern{\'a}ndez-Garc{\i}a},\ and\
\citenamefont {Piro}}]{toral2001analytical}%
\BibitemOpen
\bibfield  {author} {\bibinfo {author} {\bibfnamefont {R.}~\bibnamefont
{Toral}}, \bibinfo {author} {\bibfnamefont {C.~R.}\ \bibnamefont {Mirasso}},
\bibinfo {author} {\bibfnamefont {E.}~\bibnamefont
{Hern{\'a}ndez-Garc{\i}a}}, \ and\ \bibinfo {author} {\bibfnamefont
{O.}~\bibnamefont {Piro}},\ }\bibfield  {title} {\enquote {\bibinfo {title}
{Analytical and numerical studies of noise-induced synchronization of chaotic
systems},}\ }\href@noop {} {\bibfield  {journal} {\bibinfo  {journal} {Chaos:
An Interdisciplinary Journal of Nonlinear Science}\ }\textbf {\bibinfo
{volume} {11}},\ \bibinfo {pages} {665--673} (\bibinfo {year}
{2001})}\BibitemShut {NoStop}%
\bibitem [{\citenamefont {Matsumoto}\ and\ \citenamefont
{Tsuda}(1983)}]{matsumoto1983noise}%
\BibitemOpen
\bibfield  {author} {\bibinfo {author} {\bibfnamefont {K.}~\bibnamefont
{Matsumoto}}\ and\ \bibinfo {author} {\bibfnamefont {I.}~\bibnamefont
{Tsuda}},\ }\bibfield  {title} {\enquote {\bibinfo {title} {Noise-induced
order},}\ }\href@noop {} {\bibfield  {journal} {\bibinfo  {journal} {Journal
of Statistical Physics}\ }\textbf {\bibinfo {volume} {31}},\ \bibinfo {pages}
{87--106} (\bibinfo {year} {1983})}\BibitemShut {NoStop}%
\bibitem [{\citenamefont {Gammaitoni}\ \emph {et~al.}(1998)\citenamefont
{Gammaitoni}, \citenamefont {H{\"a}nggi}, \citenamefont {Jung},\ and\
\citenamefont {Marchesoni}}]{gammaitoni1998stochastic}%
\BibitemOpen
\bibfield  {author} {\bibinfo {author} {\bibfnamefont {L.}~\bibnamefont
{Gammaitoni}}, \bibinfo {author} {\bibfnamefont {P.}~\bibnamefont
{H{\"a}nggi}}, \bibinfo {author} {\bibfnamefont {P.}~\bibnamefont {Jung}}, \
and\ \bibinfo {author} {\bibfnamefont {F.}~\bibnamefont {Marchesoni}},\
}\bibfield  {title} {\enquote {\bibinfo {title} {Stochastic resonance},}\
}\href@noop {} {\bibfield  {journal} {\bibinfo  {journal} {Reviews of Modern
Physics}\ }\textbf {\bibinfo {volume} {70}},\ \bibinfo {pages} {223}
(\bibinfo {year} {1998})}\BibitemShut {NoStop}%
\bibitem [{\citenamefont {Benzi}\ \emph {et~al.}(1981)\citenamefont {Benzi},
\citenamefont {Sutera},\ and\ \citenamefont {Vulpiani}}]{benzi1981mechanism}%
\BibitemOpen
\bibfield  {author} {\bibinfo {author} {\bibfnamefont {R.}~\bibnamefont
{Benzi}}, \bibinfo {author} {\bibfnamefont {A.}~\bibnamefont {Sutera}}, \
and\ \bibinfo {author} {\bibfnamefont {A.}~\bibnamefont {Vulpiani}},\
}\bibfield  {title} {\enquote {\bibinfo {title} {The mechanism of stochastic
resonance},}\ }\href@noop {} {\bibfield  {journal} {\bibinfo  {journal}
{Journal of Physics A: Mathematical and General}\ }\textbf {\bibinfo {volume}
{14}},\ \bibinfo {pages} {L453} (\bibinfo {year} {1981})}\BibitemShut
{NoStop}%
\bibitem [{\citenamefont {Benzi}\ \emph {et~al.}(1982)\citenamefont {Benzi},
\citenamefont {Parisi}, \citenamefont {Sutera},\ and\ \citenamefont
{Vulpiani}}]{benzi1982stochastic}%
\BibitemOpen
\bibfield  {author} {\bibinfo {author} {\bibfnamefont {R.}~\bibnamefont
{Benzi}}, \bibinfo {author} {\bibfnamefont {G.}~\bibnamefont {Parisi}},
\bibinfo {author} {\bibfnamefont {A.}~\bibnamefont {Sutera}}, \ and\ \bibinfo
{author} {\bibfnamefont {A.}~\bibnamefont {Vulpiani}},\ }\bibfield  {title}
{\enquote {\bibinfo {title} {Stochastic resonance in climatic change},}\
}\href@noop {} {\bibfield  {journal} {\bibinfo  {journal} {Tellus}\ }\textbf
{\bibinfo {volume} {34}},\ \bibinfo {pages} {10--16} (\bibinfo {year}
{1982})}\BibitemShut {NoStop}%
\bibitem [{\citenamefont {Nicolis}\ and\ \citenamefont
{Nicolis}(1981)}]{nicolis1981stochastic}%
\BibitemOpen
\bibfield  {author} {\bibinfo {author} {\bibfnamefont {C.}~\bibnamefont
{Nicolis}}\ and\ \bibinfo {author} {\bibfnamefont {G.}~\bibnamefont
{Nicolis}},\ }\bibfield  {title} {\enquote {\bibinfo {title} {Stochastic
aspects of climatic transitions--additive fluctuations},}\ }\href@noop {}
{\bibfield  {journal} {\bibinfo  {journal} {Tellus}\ }\textbf {\bibinfo
{volume} {33}},\ \bibinfo {pages} {225--234} (\bibinfo {year}
{1981})}\BibitemShut {NoStop}%
\bibitem [{\citenamefont {Fauve}\ and\ \citenamefont
{Heslot}(1983)}]{fauve1983stochastic}%
\BibitemOpen
\bibfield  {author} {\bibinfo {author} {\bibfnamefont {S.}~\bibnamefont
{Fauve}}\ and\ \bibinfo {author} {\bibfnamefont {F.}~\bibnamefont {Heslot}},\
}\bibfield  {title} {\enquote {\bibinfo {title} {Stochastic resonance in a
bistable system},}\ }\href@noop {} {\bibfield  {journal} {\bibinfo  {journal}
{Physics Letters A}\ }\textbf {\bibinfo {volume} {97}},\ \bibinfo {pages}
{5--7} (\bibinfo {year} {1983})}\BibitemShut {NoStop}%
\bibitem [{\citenamefont {Douglass}\ \emph {et~al.}(1993)\citenamefont
{Douglass}, \citenamefont {Wilkens}, \citenamefont {Pantazelou},\ and\
\citenamefont {Moss}}]{douglass1993noise}%
\BibitemOpen
\bibfield  {author} {\bibinfo {author} {\bibfnamefont {J.~K.}\ \bibnamefont
{Douglass}}, \bibinfo {author} {\bibfnamefont {L.}~\bibnamefont {Wilkens}},
\bibinfo {author} {\bibfnamefont {E.}~\bibnamefont {Pantazelou}}, \ and\
\bibinfo {author} {\bibfnamefont {F.}~\bibnamefont {Moss}},\ }\bibfield
{title} {\enquote {\bibinfo {title} {Noise enhancement of information
transfer in crayfish mechanoreceptors by stochastic resonance},}\ }\href@noop
{} {\bibfield  {journal} {\bibinfo  {journal} {Nature}\ }\textbf {\bibinfo
{volume} {365}},\ \bibinfo {pages} {337} (\bibinfo {year}
{1993})}\BibitemShut {NoStop}%
\bibitem [{\citenamefont {Russell}\ \emph {et~al.}(1999)\citenamefont
{Russell}, \citenamefont {Wilkens},\ and\ \citenamefont
{Moss}}]{russell1999use}%
\BibitemOpen
\bibfield  {author} {\bibinfo {author} {\bibfnamefont {D.~F.}\ \bibnamefont
{Russell}}, \bibinfo {author} {\bibfnamefont {L.~A.}\ \bibnamefont
{Wilkens}}, \ and\ \bibinfo {author} {\bibfnamefont {F.}~\bibnamefont
{Moss}},\ }\bibfield  {title} {\enquote {\bibinfo {title} {Use of behavioural
stochastic resonance by paddle fish for feeding},}\ }\href@noop {} {\bibfield
{journal} {\bibinfo  {journal} {Nature}\ }\textbf {\bibinfo {volume}
{402}},\ \bibinfo {pages} {291} (\bibinfo {year} {1999})}\BibitemShut
{NoStop}%
\bibitem [{\citenamefont {L{\"o}fstedt}\ and\ \citenamefont
{Coppersmith}(1994)}]{lofstedt1994quantum}%
\BibitemOpen
\bibfield  {author} {\bibinfo {author} {\bibfnamefont {R.}~\bibnamefont
{L{\"o}fstedt}}\ and\ \bibinfo {author} {\bibfnamefont {S.N.}~\bibnamefont
{Coppersmith}},\ }\bibfield  {title} {\enquote {\bibinfo {title} {Quantum
stochastic resonance},}\ }\href@noop {} {\bibfield  {journal} {\bibinfo
{journal} {Physical Review Letters}\ }\textbf {\bibinfo {volume} {72}},\
\bibinfo {pages} {1947} (\bibinfo {year} {1994})}\BibitemShut {NoStop}%
\bibitem [{\citenamefont {Grifoni}\ and\ \citenamefont
{H{\"a}nggi}(1996)}]{grifoni1996coherent}%
\BibitemOpen
\bibfield  {author} {\bibinfo {author} {\bibfnamefont {M.}~\bibnamefont
{Grifoni}}\ and\ \bibinfo {author} {\bibfnamefont {P.}~\bibnamefont
{H{\"a}nggi}},\ }\bibfield  {title} {\enquote {\bibinfo {title} {Coherent and
incoherent quantum stochastic resonance},}\ }\href@noop {} {\bibfield
{journal} {\bibinfo  {journal} {Physical Review Letters}\ }\textbf {\bibinfo
{volume} {76}},\ \bibinfo {pages} {1611} (\bibinfo {year}
{1996})}\BibitemShut {NoStop}%
\bibitem [{\citenamefont {Grifoni}\ \emph {et~al.}(1996)\citenamefont
	{Grifoni}, \citenamefont {Hartmann}, \citenamefont {Berchtold},\ and\
	\citenamefont {H{\"a}nggi}}]{grifoni1996quantum}%
\BibitemOpen
\bibfield  {author} {\bibinfo {author} {\bibfnamefont {M.}~\bibnamefont
		{Grifoni}}, \bibinfo {author} {\bibfnamefont {L.}~\bibnamefont {Hartmann}},
	\bibinfo {author} {\bibfnamefont {S.}~\bibnamefont {Berchtold}}, \ and\
	\bibinfo {author} {\bibfnamefont {P.}~\bibnamefont {H{\"a}nggi}},\ }\bibfield
{title} {\enquote {\bibinfo {title} {Quantum tunneling and stochastic
			resonance},}\ }\href@noop {} {\bibfield  {journal} {\bibinfo  {journal}
		{Physical Review E}\ }\textbf {\bibinfo {volume} {53}},\ \bibinfo {pages}
	{5890} (\bibinfo {year} {1996})}\BibitemShut {NoStop}%
\bibitem [{\citenamefont {Wagner}\ \emph {et~al.}(2019)\citenamefont {Wagner},
\citenamefont {Talkner}, \citenamefont {Bayer}, \citenamefont {Rugeramigabo},
\citenamefont {H{\"a}nggi},\ and\ \citenamefont {Haug}}]{wagner2019quantum}%
\BibitemOpen
\bibfield  {author} {\bibinfo {author} {\bibfnamefont {T.}~\bibnamefont
{Wagner}}, \bibinfo {author} {\bibfnamefont {P.}~\bibnamefont {Talkner}},
\bibinfo {author} {\bibfnamefont {J.~C.}\ \bibnamefont {Bayer}}, \bibinfo
{author} {\bibfnamefont {E.~P.}\ \bibnamefont {Rugeramigabo}}, \bibinfo
{author} {\bibfnamefont {P.}~\bibnamefont {H{\"a}nggi}}, \ and\ \bibinfo
{author} {\bibfnamefont {R.~J.}\ \bibnamefont {Haug}},\ }\bibfield  {title}
{\enquote {\bibinfo {title} {Quantum stochastic resonance in an ac-driven
single-electron quantum dot},}\ }\href@noop {} {\bibfield  {journal}
{\bibinfo  {journal} {Nature Physics}\ }\textbf {\bibinfo {volume} {15}},\
\bibinfo {pages} {330--334} (\bibinfo {year} {2019})}\BibitemShut {NoStop}%
\bibitem [{\citenamefont {Pikovsky}\ and\ \citenamefont
{Kurths}(1997)}]{pikovsky1997coherence}%
\BibitemOpen
\bibfield  {author} {\bibinfo {author} {\bibfnamefont {A.~S.}\ \bibnamefont
{Pikovsky}}\ and\ \bibinfo {author} {\bibfnamefont {J.}~\bibnamefont
{Kurths}},\ }\bibfield  {title} {\enquote {\bibinfo {title} {Coherence
resonance in a noise-driven excitable system},}\ }\href@noop {} {\bibfield
{journal} {\bibinfo  {journal} {Physical Review Letters}\ }\textbf {\bibinfo
{volume} {78}},\ \bibinfo {pages} {775} (\bibinfo {year} {1997})}\BibitemShut
{NoStop}%
\bibitem [{\citenamefont {Gang}\ \emph {et~al.}(1993)\citenamefont {Gang},
\citenamefont {Ditzinger}, \citenamefont {Ning},\ and\ \citenamefont
{Haken}}]{gang1993stochastic}%
\BibitemOpen
\bibfield  {author} {\bibinfo {author} {\bibfnamefont {H.}~\bibnamefont
{Gang}}, \bibinfo {author} {\bibfnamefont {T.}~\bibnamefont {Ditzinger}},
\bibinfo {author} {\bibfnamefont {C.~Z.}\ \bibnamefont {Ning}}, \ and\
\bibinfo {author} {\bibfnamefont {H.}~\bibnamefont {Haken}},\ }\bibfield
{title} {\enquote {\bibinfo {title} {Stochastic resonance without external
periodic force},}\ }\href@noop {} {\bibfield  {journal} {\bibinfo  {journal}
{Physical Review Letters}\ }\textbf {\bibinfo {volume} {71}},\ \bibinfo
{pages} {807} (\bibinfo {year} {1993})}\BibitemShut {NoStop}%
\bibitem [{\citenamefont {Anishchenko}\ \emph {et~al.}(2007)\citenamefont
{Anishchenko}, \citenamefont {Astakhov}, \citenamefont {Neiman},
\citenamefont {Vadivasova},\ and\ \citenamefont
{Schimansky-Geier}}]{anishchenko2007nonlinear}%
\BibitemOpen
\bibfield  {author} {\bibinfo {author} {\bibfnamefont {V.~S.}\ \bibnamefont
{Anishchenko}}, \bibinfo {author} {\bibfnamefont {V.}~\bibnamefont
{Astakhov}}, \bibinfo {author} {\bibfnamefont {A.}~\bibnamefont {Neiman}},
\bibinfo {author} {\bibfnamefont {T.}~\bibnamefont {Vadivasova}}, \ and\
\bibinfo {author} {\bibfnamefont {L.}~\bibnamefont {Schimansky-Geier}},\
}\href@noop {} {\emph {\bibinfo {title} {Nonlinear dynamics of chaotic and
stochastic systems: tutorial and modern developments}}}\ (\bibinfo
{publisher} {Springer},\ \bibinfo {year} {2007})\BibitemShut {NoStop}%
\bibitem [{\citenamefont {Palenzuela}\ \emph {et~al.}(2001)\citenamefont
{Palenzuela}, \citenamefont {Toral}, \citenamefont {Mirasso}, \citenamefont
{Calvo},\ and\ \citenamefont {Gunton}}]{palenzuela2001coherence}%
\BibitemOpen
\bibfield  {author} {\bibinfo {author} {\bibfnamefont {C.}~\bibnamefont
{Palenzuela}}, \bibinfo {author} {\bibfnamefont {R.}~\bibnamefont {Toral}},
\bibinfo {author} {\bibfnamefont {C.~R.}\ \bibnamefont {Mirasso}}, \bibinfo
{author} {\bibfnamefont {O.}~\bibnamefont {Calvo}}, \ and\ \bibinfo {author}
{\bibfnamefont {J.~D.}\ \bibnamefont {Gunton}},\ }\bibfield  {title}
{\enquote {\bibinfo {title} {Coherence resonance in chaotic systems},}\
}\href@noop {} {\bibfield  {journal} {\bibinfo  {journal} {EPL (Europhysics
Letters)}\ }\textbf {\bibinfo {volume} {56}},\ \bibinfo {pages} {347}
(\bibinfo {year} {2001})}\BibitemShut {NoStop}%
\bibitem [{\citenamefont {Perc}(2005)}]{perc2005spatial}%
\BibitemOpen
\bibfield  {author} {\bibinfo {author} {\bibfnamefont {M.}~\bibnamefont
{Perc}},\ }\bibfield  {title} {\enquote {\bibinfo {title} {Spatial coherence
resonance in excitable media},}\ }\href@noop {} {\bibfield  {journal}
{\bibinfo  {journal} {Physical Review E}\ }\textbf {\bibinfo {volume} {72}},\
\bibinfo {pages} {016207} (\bibinfo {year} {2005})}\BibitemShut {NoStop}%
\bibitem [{\citenamefont {Mompo}\ \emph {et~al.}(2018)\citenamefont {Mompo},
\citenamefont {Ruiz-Garcia}, \citenamefont {Carretero}, \citenamefont
{Grahn}, \citenamefont {Zhang},\ and\ \citenamefont
{Bonilla}}]{mompo2018coherence}%
\BibitemOpen
\bibfield  {author} {\bibinfo {author} {\bibfnamefont {E.}~\bibnamefont
{Mompo}}, \bibinfo {author} {\bibfnamefont {M.}~\bibnamefont {Ruiz-Garcia}},
\bibinfo {author} {\bibfnamefont {M.}~\bibnamefont {Carretero}}, \bibinfo
{author} {\bibfnamefont {H.~T.}\ \bibnamefont {Grahn}}, \bibinfo {author}
{\bibfnamefont {Y.}~\bibnamefont {Zhang}}, \ and\ \bibinfo {author}
{\bibfnamefont {L.~L.}\ \bibnamefont {Bonilla}},\ }\bibfield  {title}
{\enquote {\bibinfo {title} {Coherence resonance and stochastic resonance in
an excitable semiconductor superlattice},}\ }\href@noop {} {\bibfield
{journal} {\bibinfo  {journal} {Physical Review Letters}\ }\textbf {\bibinfo
{volume} {121}},\ \bibinfo {pages} {086805} (\bibinfo {year}
{2018})}\BibitemShut {NoStop}%
\bibitem [{\citenamefont {Yu}\ \emph {et~al.}(2018)\citenamefont {Yu},
\citenamefont {Xie}, \citenamefont {Cheng},\ and\ \citenamefont
{Fan}}]{yu2018noise}%
\BibitemOpen
\bibfield  {author} {\bibinfo {author} {\bibfnamefont {D.}~\bibnamefont
{Yu}}, \bibinfo {author} {\bibfnamefont {M.}~\bibnamefont {Xie}}, \bibinfo
{author} {\bibfnamefont {Y.}~\bibnamefont {Cheng}}, \ and\ \bibinfo {author}
{\bibfnamefont {B.}~\bibnamefont {Fan}},\ }\bibfield  {title} {\enquote
{\bibinfo {title} {Noise-induced temporal regularity and signal amplification
in an optomechanical system with parametric instability},}\ }\href@noop {}
{\bibfield  {journal} {\bibinfo  {journal} {Optics express}\ }\textbf
{\bibinfo {volume} {26}},\ \bibinfo {pages} {32433--32441} (\bibinfo {year}
{2018})}\BibitemShut {NoStop}%
\bibitem [{\citenamefont {Shuai}\ and\ \citenamefont
	{Jung}(2002)}]{shuai2002optimal}%
\BibitemOpen
\bibfield  {author} {\bibinfo {author} {\bibfnamefont {J.~W.}~\bibnamefont
		{Shuai}}\ and\ \bibinfo {author} {\bibfnamefont {P.}~\bibnamefont {Jung}},\
}\bibfield  {title} {\enquote {\bibinfo {title} {Optimal intracellular
			calcium signaling},}\ }\href@noop {} {\bibfield  {journal} {\bibinfo
		{journal} {Physical Review Letters}\ }\textbf {\bibinfo {volume} {88}},\
	\bibinfo {pages} {068102} (\bibinfo {year} {2002})}\BibitemShut {NoStop}%
\bibitem [{\citenamefont {Meinhold}\ and\ \citenamefont
	{Schimansky-Geier}(2002)}]{meinhold2002analytic}%
\BibitemOpen
\bibfield  {author} {\bibinfo {author} {\bibfnamefont {L.}~\bibnamefont
		{Meinhold}}\ and\ \bibinfo {author} {\bibfnamefont {L.}~\bibnamefont
		{Schimansky-Geier}},\ }\bibfield  {title} {\enquote {\bibinfo {title}
		{Analytic description of stochastic calcium-signaling periodicity},}\
}\href@noop {} {\bibfield  {journal} {\bibinfo  {journal} {Physical Review
			E}\ }\textbf {\bibinfo {volume} {66}},\ \bibinfo {pages} {050901(R)} (\bibinfo
	{year} {2002})}\BibitemShut {NoStop}%
\bibitem [{\citenamefont {Postnov}\ \emph {et~al.}(1999)\citenamefont
{Postnov}, \citenamefont {Han}, \citenamefont {Yim},\ and\ \citenamefont
{Sosnovtseva}}]{postnov1999experimental}%
\BibitemOpen
\bibfield  {author} {\bibinfo {author} {\bibfnamefont {D.~E.}~\bibnamefont
{Postnov}}, \bibinfo {author} {\bibfnamefont {S.~K.}\ \bibnamefont {Han}},
\bibinfo {author} {\bibfnamefont {T.~G.}\ \bibnamefont {Yim}}, \ and\
\bibinfo {author} {\bibfnamefont {O.~V.}~\bibnamefont {Sosnovtseva}},\
}\bibfield  {title} {\enquote {\bibinfo {title} {Experimental observation of
coherence resonance in cascaded excitable systems},}\ }\href@noop {}
{\bibfield  {journal} {\bibinfo  {journal} {Physical Review E}\ }\textbf
{\bibinfo {volume} {59}},\ \bibinfo {pages} {R3791} (\bibinfo {year}
{1999})}\BibitemShut {NoStop}%
\bibitem [{\citenamefont {Giacomelli}\ \emph {et~al.}(2000)\citenamefont
{Giacomelli}, \citenamefont {Giudici}, \citenamefont {Balle},\ and\
\citenamefont {Tredicce}}]{giacomelli2000experimental}%
\BibitemOpen
\bibfield  {author} {\bibinfo {author} {\bibfnamefont {G.}~\bibnamefont
{Giacomelli}}, \bibinfo {author} {\bibfnamefont {M.}~\bibnamefont {Giudici}},
\bibinfo {author} {\bibfnamefont {S.}~\bibnamefont {Balle}}, \ and\ \bibinfo
{author} {\bibfnamefont {J.~R.}\ \bibnamefont {Tredicce}},\ }\bibfield
{title} {\enquote {\bibinfo {title} {Experimental evidence of coherence
resonance in an optical system},}\ }\href@noop {} {\bibfield  {journal}
{\bibinfo  {journal} {Physical Review Letters}\ }\textbf {\bibinfo {volume}
{84}},\ \bibinfo {pages} {3298} (\bibinfo {year} {2000})}\BibitemShut
{NoStop}%
\bibitem [{\citenamefont {Ushakov}\ \emph {et~al.}(2005)\citenamefont
{Ushakov}, \citenamefont {W{\"u}nsche}, \citenamefont {Henneberger},
\citenamefont {Khovanov}, \citenamefont {Schimansky-Geier},\ and\
\citenamefont {Zaks}}]{ushakov2005coherence}%
\BibitemOpen
\bibfield  {author} {\bibinfo {author} {\bibfnamefont {O.~V.}~\bibnamefont
{Ushakov}}, \bibinfo {author} {\bibfnamefont {H.~J.}\ \bibnamefont
{W{\"u}nsche}}, \bibinfo {author} {\bibfnamefont {F.}~\bibnamefont
{Henneberger}}, \bibinfo {author} {\bibfnamefont {I.A.}~\bibnamefont
{Khovanov}}, \bibinfo {author} {\bibfnamefont {L.}~\bibnamefont
{Schimansky-Geier}}, \ and\ \bibinfo {author} {\bibfnamefont
{M.}~\bibnamefont {Zaks}},\ }\bibfield  {title} {\enquote {\bibinfo {title}
{Coherence resonance near a hopf bifurcation},}\ }\href@noop {} {\bibfield
{journal} {\bibinfo  {journal} {Physical Review Letters}\ }\textbf {\bibinfo
{volume} {95}},\ \bibinfo {pages} {123903} (\bibinfo {year}
{2005})}\BibitemShut {NoStop}%
\bibitem [{\citenamefont {Zhou}\ \emph {et~al.}(2002)\citenamefont {Zhou},
\citenamefont {Jia},\ and\ \citenamefont {Ouyang}}]{zhou2002experimental}%
\BibitemOpen
\bibfield  {author} {\bibinfo {author} {\bibfnamefont {L.~Q.}~\bibnamefont
{Zhou}}, \bibinfo {author} {\bibfnamefont {X.}~\bibnamefont {Jia}}, \ and\
\bibinfo {author} {\bibfnamefont {Q.}~\bibnamefont {Ouyang}},\ }\bibfield
{title} {\enquote {\bibinfo {title} {Experimental and numerical studies of
noise-induced coherent patterns in a subexcitable system},}\ }\href@noop {}
{\bibfield  {journal} {\bibinfo  {journal} {Physical Review Letters}\
}\textbf {\bibinfo {volume} {88}},\ \bibinfo {pages} {138301} (\bibinfo
{year} {2002})}\BibitemShut {NoStop}%
\bibitem [{\citenamefont {Wilkowski}\ \emph {et~al.}(2000)\citenamefont
{Wilkowski}, \citenamefont {Ringot}, \citenamefont {Hennequin},\ and\
\citenamefont {Garreau}}]{wilkowski2000instabilities}%
\BibitemOpen
\bibfield  {author} {\bibinfo {author} {\bibfnamefont {D.}~\bibnamefont
{Wilkowski}}, \bibinfo {author} {\bibfnamefont {J.}~\bibnamefont {Ringot}},
\bibinfo {author} {\bibfnamefont {D.}~\bibnamefont {Hennequin}}, \ and\
\bibinfo {author} {\bibfnamefont {J.~C.}\ \bibnamefont {Garreau}},\
}\bibfield  {title} {\enquote {\bibinfo {title} {Instabilities in a
magneto-optical trap: Noise-induced dynamics in an atomic system},}\
}\href@noop {} {\bibfield  {journal} {\bibinfo  {journal} {Physical Review
Letters}\ }\textbf {\bibinfo {volume} {85}},\ \bibinfo {pages} {1839}
(\bibinfo {year} {2000})}\BibitemShut {NoStop}%
\bibitem [{\citenamefont {Lee}\ \emph {et~al.}(2010)\citenamefont {Lee},
\citenamefont {Choi}, \citenamefont {Han},\ and\ \citenamefont
{Strano}}]{lee2010coherence}%
\BibitemOpen
\bibfield  {author} {\bibinfo {author} {\bibfnamefont {C.~Y.}\ \bibnamefont
{Lee}}, \bibinfo {author} {\bibfnamefont {W.}~\bibnamefont {Choi}}, \bibinfo
{author} {\bibfnamefont {J.~H.}\ \bibnamefont {Han}}, \ and\ \bibinfo
{author} {\bibfnamefont {M.~S.}\ \bibnamefont {Strano}},\ }\bibfield  {title}
{\enquote {\bibinfo {title} {Coherence resonance in a single-walled carbon
nanotube ion channel},}\ }\href@noop {} {\bibfield  {journal} {\bibinfo
{journal} {Science}\ }\textbf {\bibinfo {volume} {329}},\ \bibinfo {pages}
{1320--1324} (\bibinfo {year} {2010})}\BibitemShut {NoStop}%
\bibitem [{\citenamefont {Shao}\ \emph {et~al.}(2018)\citenamefont {Shao},
\citenamefont {Yin}, \citenamefont {Song}, \citenamefont {Liu}, \citenamefont
{Li}, \citenamefont {Zhu}, \citenamefont {Biermann}, \citenamefont {Bonilla},
\citenamefont {Grahn},\ and\ \citenamefont {Zhang}}]{shao2018fast}%
\BibitemOpen
\bibfield  {author} {\bibinfo {author} {\bibfnamefont {Z.}~\bibnamefont
{Shao}}, \bibinfo {author} {\bibfnamefont {Z.}~\bibnamefont {Yin}}, \bibinfo
{author} {\bibfnamefont {H.}~\bibnamefont {Song}}, \bibinfo {author}
{\bibfnamefont {W.}~\bibnamefont {Liu}}, \bibinfo {author} {\bibfnamefont
{X.}~\bibnamefont {Li}}, \bibinfo {author} {\bibfnamefont {J.}~\bibnamefont
{Zhu}}, \bibinfo {author} {\bibfnamefont {K.}~\bibnamefont {Biermann}},
\bibinfo {author} {\bibfnamefont {L.~L.}\ \bibnamefont {Bonilla}}, \bibinfo
{author} {\bibfnamefont {H.~T.}\ \bibnamefont {Grahn}}, \ and\ \bibinfo
{author} {\bibfnamefont {Y.}~\bibnamefont {Zhang}},\ }\bibfield  {title}
{\enquote {\bibinfo {title} {Fast detection of a weak signal by a stochastic
resonance induced by a coherence resonance in an excitable gaas/al 0.45 ga
0.55 as superlattice},}\ }\href@noop {} {\bibfield  {journal} {\bibinfo
{journal} {Physical Review Letters}\ }\textbf {\bibinfo {volume} {121}},\
\bibinfo {pages} {086806} (\bibinfo {year} {2018})}\BibitemShut {NoStop}%
%
\bibitem [{\citenamefont {Hempstead}\ and\ \citenamefont
{Lax}(1967)}]{hempstead1967classical}%
\BibitemOpen
\bibfield  {author} {\bibinfo {author} {\bibfnamefont {R.~D.}\ \bibnamefont
{Hempstead}}\ and\ \bibinfo {author} {\bibfnamefont {M.}~\bibnamefont
{Lax}},\ }\bibfield  {title} {\enquote {\bibinfo {title} {Classical noise.
vi. noise in self-sustained oscillators near threshold},}\ }\href@noop {}
{\bibfield  {journal} {\bibinfo  {journal} {Physical Review}\ }\textbf
{\bibinfo {volume} {161}},\ \bibinfo {pages} {350} (\bibinfo {year}
{1967})}\BibitemShut {NoStop}%
\bibitem [{\citenamefont {Lee}\ and\ \citenamefont
{Sadeghpour}(2013)}]{lee2013quantum}%
\BibitemOpen
\bibfield  {author} {\bibinfo {author} {\bibfnamefont {T.~E.}\ \bibnamefont
{Lee}}\ and\ \bibinfo {author} {\bibfnamefont {H.~R.}~\bibnamefont
{Sadeghpour}},\ }\bibfield  {title} {\enquote {\bibinfo {title} {Quantum
synchronization of quantum van der pol oscillators with trapped ions},}\
}\href@noop {} {\bibfield  {journal} {\bibinfo  {journal} {Physical Review
Letters}\ }\textbf {\bibinfo {volume} {111}},\ \bibinfo {pages} {234101}
(\bibinfo {year} {2013})}\BibitemShut {NoStop}%
\bibitem [{\citenamefont {Walter}\ \emph {et~al.}(2014)\citenamefont {Walter},
\citenamefont {Nunnenkamp},\ and\ \citenamefont
{Bruder}}]{walter2014quantum}%
\BibitemOpen
\bibfield  {author} {\bibinfo {author} {\bibfnamefont {S.}~\bibnamefont
{Walter}}, \bibinfo {author} {\bibfnamefont {A.}~\bibnamefont {Nunnenkamp}},
\ and\ \bibinfo {author} {\bibfnamefont {C.}~\bibnamefont {Bruder}},\
}\bibfield  {title} {\enquote {\bibinfo {title} {Quantum synchronization of a
driven self-sustained oscillator},}\ }\href@noop {} {\bibfield  {journal}
{\bibinfo  {journal} {Physical Review Letters}\ }\textbf {\bibinfo {volume}
{112}},\ \bibinfo {pages} {094102} (\bibinfo {year} {2014})}\BibitemShut
{NoStop}%
\bibitem [{\citenamefont {Sonar}\ \emph {et~al.}(2018)\citenamefont {Sonar},
\citenamefont {Hajdu{\v{s}}ek}, \citenamefont {Mukherjee}, \citenamefont
{Fazio}, \citenamefont {Vedral}, \citenamefont {Vinjanampathy},\ and\
\citenamefont {Kwek}}]{sonar2018squeezing}%
\BibitemOpen
\bibfield  {author} {\bibinfo {author} {\bibfnamefont {S.}~\bibnamefont
{Sonar}}, \bibinfo {author} {\bibfnamefont {M.}~\bibnamefont
{Hajdu{\v{s}}ek}}, \bibinfo {author} {\bibfnamefont {M.}~\bibnamefont
{Mukherjee}}, \bibinfo {author} {\bibfnamefont {R.}~\bibnamefont {Fazio}},
\bibinfo {author} {\bibfnamefont {V.}~\bibnamefont {Vedral}}, \bibinfo
{author} {\bibfnamefont {S.}~\bibnamefont {Vinjanampathy}}, \ and\ \bibinfo
{author} {\bibfnamefont {L.-C.}\ \bibnamefont {Kwek}},\ }\bibfield  {title}
{\enquote {\bibinfo {title} {Squeezing enhances quantum synchronization},}\
}\href@noop {} {\bibfield  {journal} {\bibinfo  {journal} {Physical Review
Letters}\ }\textbf {\bibinfo {volume} {120}},\ \bibinfo {pages} {163601}
(\bibinfo {year} {2018})}\BibitemShut {NoStop}%
%
\bibitem [{\citenamefont {Kato}\ \emph {et~al.}(2019)\citenamefont {Kato},
	\citenamefont {Yamamoto},\ and\ \citenamefont
	{Nakao}}]{kato2019semiclassical}%
\BibitemOpen
\bibfield  {author} {\bibinfo {author} {\bibfnamefont {Y.}~\bibnamefont
		{Kato}}, \bibinfo {author} {\bibfnamefont {N.}~\bibnamefont {Yamamoto}}, \
	and\ \bibinfo {author} {\bibfnamefont {H.}~\bibnamefont {Nakao}},\ }\href
{\doibase 10.1103/PhysRevResearch.1.033012} {\bibfield  {journal} {\bibinfo
		{journal} {Phys. Rev. Research}\ }\textbf {\bibinfo {volume} {1}},\ \bibinfo
	{pages} {033012} (\bibinfo {year} {2019})}\BibitemShut {NoStop}%
%
\bibitem [{\citenamefont {Weiss}\ \emph {et~al.}(2017)\citenamefont {Weiss},
\citenamefont {Walter},\ and\ \citenamefont {Marquardt}}]{weiss2017quantum}%
\BibitemOpen
\bibfield  {author} {\bibinfo {author} {\bibfnamefont {T.}~\bibnamefont
{Weiss}}, \bibinfo {author} {\bibfnamefont {S.}~\bibnamefont {Walter}}, \
and\ \bibinfo {author} {\bibfnamefont {F.}~\bibnamefont {Marquardt}},\
}\bibfield  {title} {\enquote {\bibinfo {title} {Quantum-coherent phase
oscillations in synchronization},}\ }\href@noop {} {\bibfield  {journal}
{\bibinfo  {journal} {Physical Review A}\ }\textbf {\bibinfo {volume} {95}},\
\bibinfo {pages} {041802(R)} (\bibinfo {year} {2017})}\BibitemShut {NoStop}%
%
\bibitem [{\citenamefont {L{\"o}rch}\ \emph {et~al.}(2016)\citenamefont
{L{\"o}rch}, \citenamefont {Amitai}, \citenamefont {Nunnenkamp},\ and\
\citenamefont {Bruder}}]{lorch2016genuine}%
\BibitemOpen
\bibfield  {author} {\bibinfo {author} {\bibfnamefont {N.}~\bibnamefont
{L{\"o}rch}}, \bibinfo {author} {\bibfnamefont {E.}~\bibnamefont {Amitai}},
\bibinfo {author} {\bibfnamefont {A.}~\bibnamefont {Nunnenkamp}}, \ and\
\bibinfo {author} {\bibfnamefont {C.}~\bibnamefont {Bruder}},\ }\bibfield
{title} {\enquote {\bibinfo {title} {Genuine quantum signatures in
synchronization of anharmonic self-oscillators},}\ }\href@noop {} {\bibfield
{journal} {\bibinfo  {journal} {Physical Review Letters}\ }\textbf {\bibinfo
{volume} {117}},\ \bibinfo {pages} {073601} (\bibinfo {year}
{2016})}\BibitemShut {NoStop}%
\bibitem [{\citenamefont {Kato}\ and\ \citenamefont
	{Nakao}(2020)}]{kato2020semiclassical}%
\BibitemOpen
\bibfield  {author} {\bibinfo {author} {\bibfnamefont {Y.}~\bibnamefont
		{Kato}}\ and\ \bibinfo {author} {\bibfnamefont {H.}~\bibnamefont {Nakao}},\
}\bibfield  {title} {\enquote {\bibinfo {title} {Semiclassical optimization
			of entrainment stability and phase coherence in weakly forced quantum
			limit-cycle oscillators},}\ }\href@noop {} {\bibfield  {journal} {\bibinfo
		{journal} {Physical Review E}\ }\textbf {\bibinfo {volume} {101}},\ \bibinfo
	{pages} {012210} (\bibinfo {year} {2020})}\BibitemShut {NoStop}%
\bibitem [{\citenamefont {Bastidas}\ \emph {et~al.}(2015)\citenamefont
	{Bastidas}, \citenamefont {Omelchenko}, \citenamefont {Zakharova},
	\citenamefont {Sch{\"o}ll},\ and\ \citenamefont
	{Brandes}}]{bastidas2015quantum}%
\BibitemOpen
\bibfield  {author} {\bibinfo {author} {\bibfnamefont {V.}~\bibnamefont
		{Bastidas}}, \bibinfo {author} {\bibfnamefont {I.}~\bibnamefont
		{Omelchenko}}, \bibinfo {author} {\bibfnamefont {A.}~\bibnamefont
		{Zakharova}}, \bibinfo {author} {\bibfnamefont {E.}~\bibnamefont
		{Sch{\"o}ll}}, \ and\ \bibinfo {author} {\bibfnamefont {T.}~\bibnamefont
		{Brandes}},\ }\bibfield  {title} {\enquote {\bibinfo {title} {Quantum
			signatures of chimera states},}\ }\href@noop {} {\bibfield  {journal}
	{\bibinfo  {journal} {Physical Review E}\ }\textbf {\bibinfo {volume} {92}},\
	\bibinfo {pages} {062924} (\bibinfo {year} {2015})}\BibitemShut {NoStop}%
\bibitem [{\citenamefont {Es'~haqi Sani}\ \emph {et~al.}(2020)\citenamefont
	{Es'~haqi Sani}, \citenamefont {Manzano}, \citenamefont {Zambrini},\ and\
	\citenamefont {Fazio}}]{es2020synchronization}%
\BibitemOpen
\bibfield  {author} {\bibinfo {author} {\bibfnamefont {N.}~\bibnamefont
		{Es'~haqi Sani}}, \bibinfo {author} {\bibfnamefont {G.}~\bibnamefont
		{Manzano}}, \bibinfo {author} {\bibfnamefont {R.}~\bibnamefont {Zambrini}}, \
	and\ \bibinfo {author} {\bibfnamefont {R.}~\bibnamefont {Fazio}},\ }\bibfield
{title} {\enquote {\bibinfo {title} {Synchronization along quantum
			trajectories},}\ }\href@noop {} {\bibfield  {journal} {\bibinfo  {journal}
		{Physical Review Research}\ }\textbf {\bibinfo {volume} {2}},\ \bibinfo
	{pages} {023101} (\bibinfo {year} {2020})}\BibitemShut {NoStop}%
\bibitem [{\citenamefont {Mok}\ \emph {et~al.}(2020)\citenamefont {Mok},
	\citenamefont {Kwek},\ and\ \citenamefont
	{Heimonen}}]{mok2020synchronization}%
\BibitemOpen
\bibfield  {author} {\bibinfo {author} {\bibfnamefont {W.-K.}\ \bibnamefont
		{Mok}}, \bibinfo {author} {\bibfnamefont {L.-C.}\ \bibnamefont {Kwek}}, \
	and\ \bibinfo {author} {\bibfnamefont {H.}~\bibnamefont {Heimonen}},\
}\bibfield  {title} {\enquote {\bibinfo {title} {Synchronization boost with
			single-photon dissipation in the deep quantum regime},}\ }\href@noop {}
{\bibfield  {journal} {\bibinfo  {journal} {Physical Review Research}\
	}\textbf {\bibinfo {volume} {2}},\ \bibinfo {pages} {033422} (\bibinfo {year}
	{2020})}\BibitemShut {NoStop}%
\bibitem [{\citenamefont {Bandyopadhyay}\ \emph {et~al.}(2020)\citenamefont
	{Bandyopadhyay}, \citenamefont {Khatun}, \citenamefont {Biswas},\ and\
	\citenamefont {Banerjee}}]{bandyopadhyay2020quantum}%
\BibitemOpen
\bibfield  {author} {\bibinfo {author} {\bibfnamefont {B.}~\bibnamefont
		{Bandyopadhyay}}, \bibinfo {author} {\bibfnamefont {T.}~\bibnamefont
		{Khatun}}, \bibinfo {author} {\bibfnamefont {D.}~\bibnamefont {Biswas}}, \
	and\ \bibinfo {author} {\bibfnamefont {T.}~\bibnamefont {Banerjee}},\
}\href@noop {} {\bibfield  {journal} {\bibinfo  {journal} {Physical Review
			E}\ }\textbf {\bibinfo {volume} {102}},\ \bibinfo {pages} {062205} (\bibinfo
	{year} {2020})}\BibitemShut {NoStop}%
\bibitem [{\citenamefont {Strogatz}(1994)}]{strogatz1994nonlinear}%
\BibitemOpen
\bibfield  {author} {\bibinfo {author} {\bibfnamefont {S.}~\bibnamefont
{Strogatz}},\ }\href@noop {} {\emph {\bibinfo {title} {Nonlinear dynamics and
chaos}}}\ (\bibinfo  {publisher} {Westview Press},\ \bibinfo {year}
{1994})\BibitemShut {NoStop}%
\bibitem [{\citenamefont {Guckenheimer}\ and\ \citenamefont
{Holmes}(1983)}]{guckenheimer1983nonlinear}%
\BibitemOpen
\bibfield  {author} {\bibinfo {author} {\bibfnamefont {J.}~\bibnamefont
{Guckenheimer}}\ and\ \bibinfo {author} {\bibfnamefont {P.}~\bibnamefont
{Holmes}},\ }\href@noop {} {\emph {\bibinfo {title} {Nonlinear Oscillations,
Dynamical Systems, and Bifurcations of Vector Fields}}}\ (\bibinfo
{publisher} {Springer},\ \bibinfo {year} {1983})\BibitemShut {NoStop}%
\bibitem [{\citenamefont {Gardiner}\ and\ \citenamefont
	{Haken}(1991)}]{gardiner1991quantum}%
\BibitemOpen
\bibfield  {author} {\bibinfo {author} {\bibfnamefont {C.~W.}\ \bibnamefont
		{Gardiner}}\ and\ \bibinfo {author} {\bibfnamefont {H.}~\bibnamefont
		{Haken}},\ }\href@noop {} {\emph {\bibinfo {title} {Quantum noise}}}\
(\bibinfo  {publisher} {Springer},\ \bibinfo {year} {1991})\BibitemShut
{NoStop}%
\bibitem [{\citenamefont {Van Der~Pol}(1927)}]{van1927vii}%
\BibitemOpen
\bibfield  {author} {\bibinfo {author} {\bibfnamefont {B.}~\bibnamefont {Van
			Der~Pol}},\ }\href@noop {} {\bibfield  {journal} {\bibinfo  {journal} {The
			London, Edinburgh, and Dublin Philosophical Magazine and Journal of Science}\
	}\textbf {\bibinfo {volume} {3}},\ \bibinfo {pages} {65} (\bibinfo {year}
	{1927})}\BibitemShut {NoStop}%
\bibitem [{\citenamefont {Carmichael}(2007)}]{carmichael2007statistical}%
\BibitemOpen
\bibfield  {author} {\bibinfo {author} {\bibfnamefont {H.~J.}\ \bibnamefont
{Carmichael}},\ }\href@noop {} {\emph {\bibinfo {title} {Statistical Methods
in Quantum Optics 1, 2}}}\ (\bibinfo  {publisher} {Springer},\ \bibinfo
{year} {2007})\BibitemShut {NoStop}%
\bibitem [{\citenamefont {Kuramoto}(1984)}]{kuramoto1984chemical}%
\BibitemOpen
\bibfield  {author} {\bibinfo {author} {\bibfnamefont {Y.}~\bibnamefont
		{Kuramoto}},\ }\href@noop {} {\emph {\bibinfo {title} {Chemical oscillations,
			waves, and turbulence}}}\ (\bibinfo  {publisher} {Springer},\ \bibinfo
{address} {Berlin},\ \bibinfo {year} {1984})\BibitemShut {NoStop}%
\bibitem [{\citenamefont {Chia}\ \emph {et~al.}(2020)\citenamefont {Chia},
	\citenamefont {Kwek},\ and\ \citenamefont {Noh}}]{chia2020relaxation}%
\BibitemOpen
\bibfield  {author} {\bibinfo {author} {\bibfnamefont {A.}~\bibnamefont
		{Chia}}, \bibinfo {author} {\bibfnamefont {L.-C.}~\bibnamefont {Kwek}}, \ and\
	\bibinfo {author} {\bibfnamefont {C.}~\bibnamefont {Noh}},\ }\bibfield
{title} {\enquote {\bibinfo {title} {Relaxation oscillations and frequency
			entrainment in quantum mechanics},}\ }\href@noop {} {\bibfield  {journal}
	{\bibinfo  {journal} {Physical Review E}\ }\textbf {\bibinfo {volume}
		{102}},\ \bibinfo {pages} {042213} (\bibinfo {year} {2020})}\BibitemShut
{NoStop}%
\bibitem [{\citenamefont {C.~W.~Wachtler}\ and\ \citenamefont
	{Munro}(2020)}]{wachtler2020dissipative}%
\BibitemOpen
\bibfield  {author} {\bibinfo {author} {\bibfnamefont {G.~S.}\ \bibnamefont
		{C.~W.~Wachtler}, \bibfnamefont {V.~M.~Bastidas}}\ and\ \bibinfo {author}
	{\bibfnamefont {W.~J.}\ \bibnamefont {Munro}},\ }\bibfield  {title} {\enquote
	{\bibinfo {title} {Dissipative nonequilibrium synchronization of topological
			edge states via self-oscillation},}\ }\href@noop {} {\bibfield  {journal}
	{\bibinfo  {journal} {Physical Review B}\ }\textbf {\bibinfo {volume}
		{102}},\ \bibinfo {pages} {014309} (\bibinfo {year} {2020})}\BibitemShut
{NoStop}%
\bibitem [{\citenamefont {Johansson}\ \emph {et~al.}(2012)\citenamefont
{Johansson}, \citenamefont {Nation},\ and\ \citenamefont
{Nori}}]{johansson2012qutip}%
\BibitemOpen
\bibfield  {author} {\bibinfo {author} {\bibfnamefont {J.}~\bibnamefont
{Johansson}}, \bibinfo {author} {\bibfnamefont {P.}~\bibnamefont {Nation}}, \
and\ \bibinfo {author} {\bibfnamefont {F.}~\bibnamefont {Nori}},\ }\bibfield
{title} {\enquote {\bibinfo {title} {Qutip: An open-source python framework
for the dynamics of open quantum systems},}\ }\href@noop {} {\bibfield
{journal} {\bibinfo  {journal} {Computer Physics Communications}\ }\textbf
{\bibinfo {volume} {183}},\ \bibinfo {pages} {1760--1772} (\bibinfo {year}
{2012})}\BibitemShut {NoStop}%
\bibitem [{\citenamefont {Johansson}\ \emph {et~al.}(2013)\citenamefont
{Johansson}, \citenamefont {Nation},\ and\ \citenamefont
{Nori}}]{johansson2013qutip}%
\BibitemOpen
\bibfield  {author} {\bibinfo {author} {\bibfnamefont {J.}~\bibnamefont
{Johansson}}, \bibinfo {author} {\bibfnamefont {P.}~\bibnamefont {Nation}}, \
and\ \bibinfo {author} {\bibfnamefont {F.}~\bibnamefont {Nori}},\ }\bibfield
{title} {\enquote {\bibinfo {title} {Qutip 2: A python framework for the
dynamics of open quantum systems},}\ }\href@noop {} {\bibfield  {journal}
{\bibinfo  {journal} {Computer Physics Communications}\ }\textbf {\bibinfo
{volume} {184}},\ \bibinfo {pages} {1234--1240} (\bibinfo {year}
{2013})}\BibitemShut {NoStop}%
\bibitem [{\citenamefont {Lindner}\ and\ \citenamefont
{Schimansky-Geier}(2000)}]{lindner2000coherence}%
\BibitemOpen
\bibfield  {author} {\bibinfo {author} {\bibfnamefont {B.}~\bibnamefont
{Lindner}}\ and\ \bibinfo {author} {\bibfnamefont {L.}~\bibnamefont
{Schimansky-Geier}},\ }\bibfield  {title} {\enquote {\bibinfo {title}
{Coherence and stochastic resonance in a two-state system},}\ }\href@noop {}
{\bibfield  {journal} {\bibinfo  {journal} {Physical Review E}\ }\textbf
{\bibinfo {volume} {61}},\ \bibinfo {pages} {6103} (\bibinfo {year}
{2000})}\BibitemShut {NoStop}%
\bibitem [{\citenamefont {Lindner}(2002)}]{lindner2002coherence}%
\BibitemOpen
\bibfield  {author} {\bibinfo {author} {\bibfnamefont {B.}~\bibnamefont
{Lindner}},\ }\href@noop {} {\emph {\bibinfo {title} {Coherence and
stochastic resonance in nonlinear dynamical systems}}}\ (\bibinfo
{publisher} {Logos-Verlag},\ \bibinfo {year} {2002})\BibitemShut {NoStop}%
\bibitem [{\citenamefont {Risken}\ \emph {et~al.}(1987)\citenamefont {Risken},
	\citenamefont {Savage}, \citenamefont {Haake},\ and\ \citenamefont
	{Walls}}]{risken1987quantum}%
\BibitemOpen
\bibfield  {author} {\bibinfo {author} {\bibfnamefont {H.}~\bibnamefont
		{Risken}}, \bibinfo {author} {\bibfnamefont {C.}~\bibnamefont {Savage}},
	\bibinfo {author} {\bibfnamefont {F.}~\bibnamefont {Haake}}, \ and\ \bibinfo
	{author} {\bibfnamefont {D.}~\bibnamefont {Walls}},\ }\bibfield  {title}
{\enquote {\bibinfo {title} {Quantum tunneling in dispersive optical
			bistability},}\ }\href@noop {} {\bibfield  {journal} {\bibinfo  {journal}
		{Physical Review A}\ }\textbf {\bibinfo {volume} {35}},\ \bibinfo {pages}
	{1729} (\bibinfo {year} {1987})}\BibitemShut {NoStop}%
\bibitem [{\citenamefont {Risken}\ and\ \citenamefont
	{Vogel}(1988)}]{risken1988quantum}%
\BibitemOpen
\bibfield  {author} {\bibinfo {author} {\bibfnamefont {H.}~\bibnamefont
		{Risken}}\ and\ \bibinfo {author} {\bibfnamefont {K.}~\bibnamefont {Vogel}},\
}\bibfield  {title} {\enquote {\bibinfo {title} {Quantum tunneling rates in
			dispersive optical bistability for low cavity damping},}\ }\href@noop {}
{\bibfield  {journal} {\bibinfo  {journal} {Physical Review A}\ }\textbf
	{\bibinfo {volume} {38}},\ \bibinfo {pages} {1349} (\bibinfo {year}
	{1988})}\BibitemShut {NoStop}%
\bibitem [{\citenamefont {Vogel}\ and\ \citenamefont
	{Risken}(1988)}]{vogel1988quantum}%
\BibitemOpen
\bibfield  {author} {\bibinfo {author} {\bibfnamefont {K.}~\bibnamefont
		{Vogel}}\ and\ \bibinfo {author} {\bibfnamefont {H.}~\bibnamefont {Risken}},\
}\bibfield  {title} {\enquote {\bibinfo {title} {Quantum-tunneling rates and
			stationary solutions in dispersive optical bistability},}\ }\href@noop {}
{\bibfield  {journal} {\bibinfo  {journal} {Physical Review A}\ }\textbf
	{\bibinfo {volume} {38}},\ \bibinfo {pages} {2409} (\bibinfo {year}
	{1988})}\BibitemShut {NoStop}%
\bibitem [{\citenamefont {Rugar}\ and\ \citenamefont
	{Gr{\"u}tter}(1991)}]{rugar1991mechanical}%
\BibitemOpen
\bibfield  {author} {\bibinfo {author} {\bibfnamefont {D.}~\bibnamefont
		{Rugar}}\ and\ \bibinfo {author} {\bibfnamefont {P.}~\bibnamefont
		{Gr{\"u}tter}},\ }\href@noop {} {\bibfield  {journal} {\bibinfo  {journal}
		{Physical Review Letters}\ }\textbf {\bibinfo {volume} {67}},\ \bibinfo
	{pages} {699} (\bibinfo {year} {1991})}\BibitemShut {NoStop}
\bibitem [{\citenamefont {Han}\ \emph {et~al.}(1999)\citenamefont {Han},
	\citenamefont {Yim}, \citenamefont {Postnov},\ and\ \citenamefont
	{Sosnovtseva}}]{han1999interacting}%
\BibitemOpen
\bibfield  {author} {\bibinfo {author} {\bibfnamefont {S.~K.}\ \bibnamefont
		{Han}}, \bibinfo {author} {\bibfnamefont {T.~G.}\ \bibnamefont {Yim}},
	\bibinfo {author} {\bibfnamefont {D.~E.}~\bibnamefont {Postnov}}, \ and\
	\bibinfo {author} {\bibfnamefont {O.~V.}~\bibnamefont {Sosnovtseva}},\
}\bibfield  {title} {\enquote {\bibinfo {title} {Interacting coherence
			resonance oscillators},}\ }\href@noop {} {\bibfield  {journal} {\bibinfo
		{journal} {Physical Review Letters}\ }\textbf {\bibinfo {volume} {83}},\
	\bibinfo {pages} {1771} (\bibinfo {year} {1999})}\BibitemShut {NoStop}%
\bibitem [{\citenamefont {Neiman}\ \emph {et~al.}(1999)\citenamefont {Neiman},
\citenamefont {Schimansky-Geier}, \citenamefont {Cornell-Bell},\ and\
\citenamefont {Moss}}]{neiman1999noise}%
\BibitemOpen
\bibfield  {author} {\bibinfo {author} {\bibfnamefont {A.}~\bibnamefont
{Neiman}}, \bibinfo {author} {\bibfnamefont {L.}~\bibnamefont
{Schimansky-Geier}}, \bibinfo {author} {\bibfnamefont {A.}~\bibnamefont
{Cornell-Bell}}, \ and\ \bibinfo {author} {\bibfnamefont {F.}~\bibnamefont
{Moss}},\ }\bibfield  {title} {\enquote {\bibinfo {title} {Noise-enhanced
phase synchronization in excitable media},}\ }\href@noop {} {\bibfield
{journal} {\bibinfo  {journal} {Physical Review Letters}\ }\textbf {\bibinfo
{volume} {83}},\ \bibinfo {pages} {4896} (\bibinfo {year}
{1999})}\BibitemShut {NoStop}%
\bibitem [{\citenamefont {Toral}\ \emph {et~al.}(2003)\citenamefont {Toral},
\citenamefont {Mirasso},\ and\ \citenamefont {Gunton}}]{toral2003system}%
\BibitemOpen
\bibfield  {author} {\bibinfo {author} {\bibfnamefont {R.}~\bibnamefont
{Toral}}, \bibinfo {author} {\bibfnamefont {C.}~\bibnamefont {Mirasso}}, \
and\ \bibinfo {author} {\bibfnamefont {J.}~\bibnamefont {Gunton}},\
}\bibfield  {title} {\enquote {\bibinfo {title} {System size coherence
resonance in coupled fitzhugh-nagumo models},}\ }\href@noop {} {\bibfield
{journal} {\bibinfo  {journal} {EPL (Europhysics Letters)}\ }\textbf
{\bibinfo {volume} {61}},\ \bibinfo {pages} {162} (\bibinfo {year}
{2003})}\BibitemShut {NoStop}%
\bibitem [{\citenamefont {Andreev}\ \emph {et~al.}(2018)\citenamefont
{Andreev}, \citenamefont {Makarov}, \citenamefont {Runnova}, \citenamefont
{Pisarchik},\ and\ \citenamefont {Hramov}}]{andreev2018coherence}%
\BibitemOpen
\bibfield  {author} {\bibinfo {author} {\bibfnamefont {A.~V.}\ \bibnamefont
{Andreev}}, \bibinfo {author} {\bibfnamefont {V.~V.}\ \bibnamefont
{Makarov}}, \bibinfo {author} {\bibfnamefont {A.~E.}\ \bibnamefont
{Runnova}}, \bibinfo {author} {\bibfnamefont {A.~N.}\ \bibnamefont
{Pisarchik}}, \ and\ \bibinfo {author} {\bibfnamefont {A.~E.}\ \bibnamefont
{Hramov}},\ }\bibfield  {title} {\enquote {\bibinfo {title} {Coherence
resonance in stimulated neuronal network},}\ }\href@noop {} {\bibfield
{journal} {\bibinfo  {journal} {Chaos, Solitons \& Fractals}\ }\textbf
{\bibinfo {volume} {106}},\ \bibinfo {pages} {80--85} (\bibinfo {year}
{2018})}\BibitemShut {NoStop}%
\end{thebibliography}
\end{document}